\documentclass[aps,pra,reprint,longbibliography,groupedaddress]{revtex4-1}

\makeatletter 
\def\@fnsymbol#1{\ensuremath{\ifcase#1\or \dagger\or *\or \ddagger\or
   \mathsection\or \mathparagraph\or \|\or **\or \dagger\dagger
   \or \ddagger\ddagger \else\@ctrerr\fi}}
\makeatother

\usepackage{amssymb,amsfonts,amsmath}
\usepackage{graphicx} 
\usepackage{bm}       
\usepackage{times}

\begin{document}

\title{Identification of molecular quantum states using phase-sensitive forces}

\author{Kaveh Najafian}
\thanks{These two authors contributed equally}
\author{Ziv Meir}
\thanks{These two authors contributed equally}
\author{Mudit Sinhal}
\author{Stefan Willitsch}
\email[To whom correspondence should be addressed: ]{stefan.willitsch@unibas.ch}

\affiliation{Department of Chemistry, University of Basel, Klingelbergstrasse 80, 4056 Basel, Switzerland}

\date{\today}

\begin{abstract}
Quantum-logic techniques used to manipulate quantum systems are now increasingly being applied to molecules.
Previous experiments on single trapped diatomic species have enabled state detection with excellent fidelities and highly precise spectroscopic measurements. 
However, for complex molecules with a dense energy-level structure improved methods are necessary.
Here, we demonstrate an enhanced quantum protocol for molecular state detection using state-dependent forces. Our approach is based on interfering a reference and a signal force applied to a single atomic and molecular ion, respectively, in order to extract their relative phase.
We use this phase information to identify states embedded in a dense molecular energy-level structure and to monitor state-to-state inelastic scattering processes. This method can also be used to exclude a large number of states in a single measurement when the initial state preparation is imperfect and information on the molecular properties is incomplete.
While the present experiments focus on N$_2^+$, the method is general and is expected to be of particular benefit for polyatomic systems.
\end{abstract}
\maketitle

\subsection*{Introduction}
The quantum control of isolated particles forms the basis for recent advancements in quantum computation \cite{wright19,friis18,arute19}, for precise time and frequency measurements \cite{brewer19,huntemann16a}, for searches for new physics beyond the standard model \cite{safronova18a,demille17a}, for controlled chemistry \cite{liu18,sikorsky18a,dorfler19b} and for quantum communication \cite{kimble08a,northup14}.

Recent progress in the quantum control of single trapped molecules \cite{wolf16a, chou17a, sinhal20, chou20a, lin19} has enabled the detection \cite{wolf16a,sinhal20}, coherent manipulation \cite{chou17a,chou20a} and entanglement \cite{lin19} of molecular quantum states with high fidelity on the single-particle level. These advances aim at encoding qubits in rotational and vibrational molecular energy states \cite{meir19a,najafian20b} and at performing precise spectroscopic measurements \cite{biesheuvel16a} with applications including the development of mid-infra-red frequency standards \cite{schiller14a} and testing a possible variation of fundamental constants such as the proton-to-electron-mass ratio \cite{schiller05a,jansen14a,flambaum07a,kajita14a,kajita16a}.

Previous works \cite{wolf16a, chou17a, sinhal20, chou20a, lin19} built on prior knowledge of the molecular energy-level structure to engineer level subspaces that enable a precise control on the quantum level. These approaches relied on the entanglement of the internal molecular states with the external motion of a molecular-ion (MI) atomic-ion (AI) Coulomb crystal \cite{leibfried12a}. The motion was generated by, e.g., applying state-dependent forces on the MI \cite{wolf16a, sinhal20}. The molecular state was then inferred from the excitation \emph{amplitude} of the MI-AI crystal read out on the AI.

However, for larger and more complex molecules such prior knowledge of the molecular energy-level structure is often not available. Moreover, situations occur, even in comparatively simple diatomic molecules, in which several states that are potentially populated result in similar excitation amplitudes generated by the state-dependent forces such that it is not readily possible to unambiguously distinguish them. Indeed, it can be expected that this is a typical scenario given that molecular-state preparation down to the hyperfine or Zeeman level is currently only feasible for the simplest systems \cite{bressel12a, chou17a}.

Here, we enhanced the capabilities of a recently developed method for the quantum-non-demolition readout of molecular states \cite{sinhal20, meir19a} by detecting the \emph{phase} of the state-dependent forces in addition to their amplitude. We achieved this by interfering a reference force applied on the AI with the state-dependent force experienced by the MI. By changing the relative phase of the interference, we were able to detect the absolute phase of the force applied on the MI in addition to its amplitude. 

Using this method, we experimentally identified specific Zeeman states of a single $^{14}$N$_2^+$ MI in different excited rotational states of its electronic and vibrational ground state. We also showed how this method can be used to track the initial and final states during state-changing processes like chemical reactions and inelastic collisions. Finally, we demonstrated a protocol to simultaneously exclude a large subset of rotational levels. This protocol is useful when probing an unknown molecular state within a complex and only partially known energy level structure.

\subsection*{Phase-sensitive forces}
\begin{figure}
	\centering
	\includegraphics[width=\linewidth,clip]{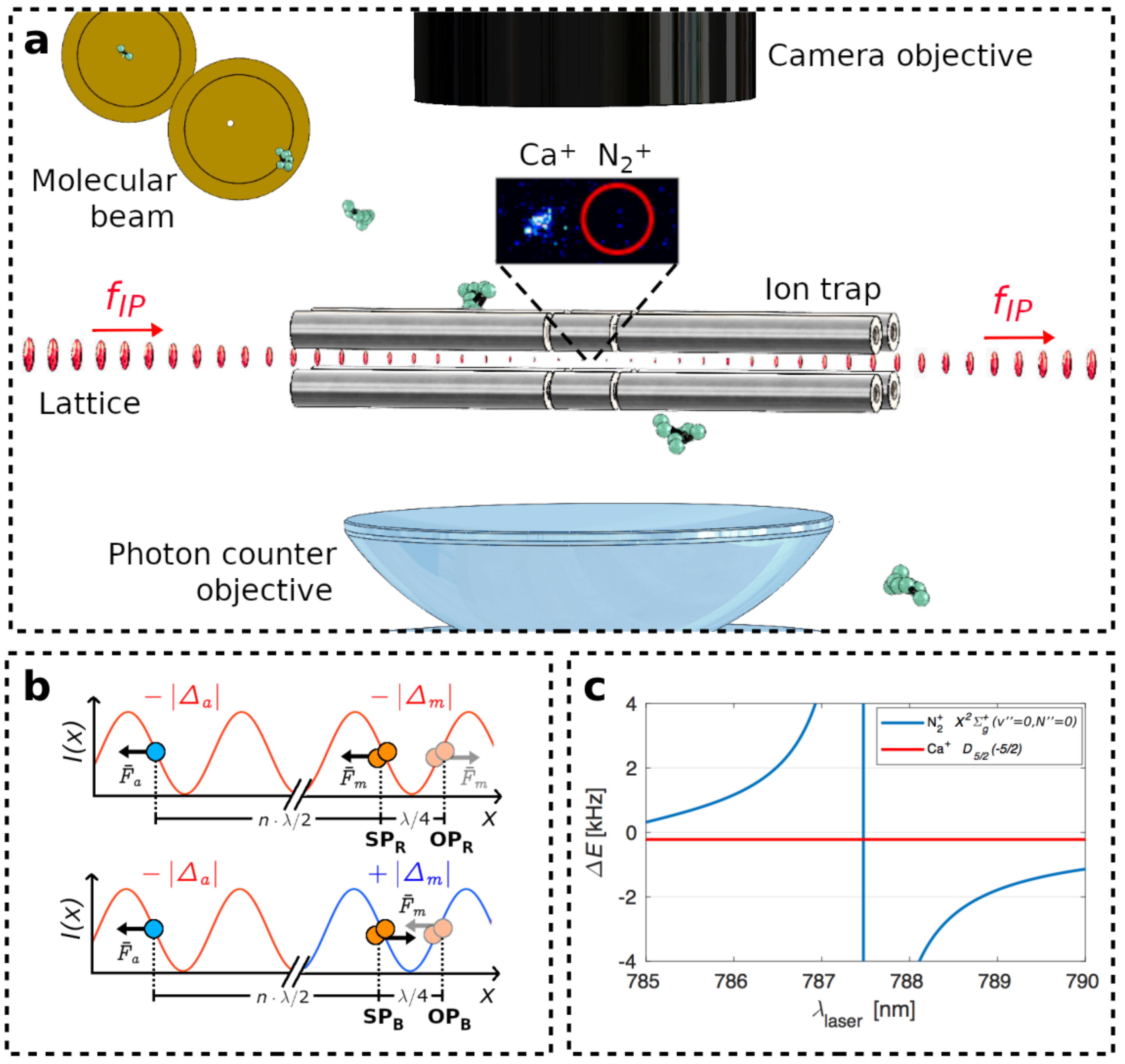}
 	\caption{\textbf{a)} Experimental setup. A pulsed molecular beam of neutral N$_2$ molecules traversed the center of a linear radiofrequency ion trap. A single $^{14}$N$_2^+$ molecular ion was generated from the molecular beam by photoionization and was trapped together with a single laser-cooled $^{40}$Ca$^+$ ion (inset). A running-wave optical lattice generated from two counter-propagating laser beams with frequency difference $\Delta f=f_{IP}$ was overlapped with the two-ion crystal to excite motion depending on the molecular state. The Ca$^+$ ion enabled sympathetic cooling of the N$_2^+$ ion as well as readout of the common motional state and provided a reference for the phase-dependent method presented here. \textbf{b)} Four possible configurations (SP$_\textrm{R/B}$, OP$_\textrm{R/B}$) of the lattice containing the atomic ion (AI, blue circle) and the molecular ion (MI, orange circles). The direction of the force on the MI (AI) ($F_{m(a)}$) was determined by the frequency detuning $\pm |\Delta_{m(a)}|$ of the lattice laser beams from resonance and the relative positions of the particles in the lattice field. The AI and the MI had either the same signs of detuning (top) or opposite signs (bottom). For a MI at positions SP$_\textrm{R}$ and SP$_\textrm{B}$ in the lattice, the distance between the two ions was such that they experienced the same phase (SP), while at positions OP$_\textrm{R}$ and OP$_\textrm{B}$ the ions experienced opposite phases (OP) of the lattice field. \textbf{c)} The ac-Stark shift, $\Delta E$, experienced by the AI in the $(3d)~^2D_{5/2}(m=-5/2)$ (red) and the MI in the $X^2\Sigma_g^+(v''=0,N''=0)$ (blue) states as a function of lattice wavelength in the region 785-790~nm. The direction of the ac-Stark shift switches sign when changing from negative to positive detuning across a molecular resonance while the ac-Stark shift experienced by the AI remains constant in this wavelength range.}
	\label{fig:cartoon}
\end{figure}

We consider a string of two ions, an AI and a MI confined in a linear radiofrequency ion trap (Figure \ref{fig:cartoon}a). The two-ion string exhibits two fundamental motional modes along the trap axis, an in-phase (IP) mode and an out-of-phase mode \cite{morigi01a,home13a}. We superimpose the ions with a running-wave optical lattice generated by two overlapping linearly polarized counterpropagating laser beams with frequency difference $\Delta f$. The lattice induces a modulated ac-Stark shift on the ions which results in a state-dependent optical-dipole force (ODF) oscillating at the frequency $\Delta f$. When the oscillation frequency of the ODF matches the one of a motional mode (e.g., $\Delta f = f_{IP}$ for the in-phase mode), the ODF resonantly excites motion in the mode \cite{wineland98a,meir19a,sinhal20}. This motion can be detected on the AI through sideband thermometry \cite{meekhof96a,leibfried03a,sinhal20}. 

In addition to the ac-Stark shift on the MI, the lattice may induce a non-negligible ac-Stark shift on the AI.. Since two ODFs are applied on the crystal simultaneously, the relative phase of the forces will influence the total degree of motional excitation. This phase depends on both the sign of the detuning of the lattice wavelength from spectroscopic resonances in the ions as well as the distance of the ions from each other. The former determines the sign of the ac-Stark shift experienced by the AI and the MI while the latter determines the relative phase of the optical lattice at the positions of the AI and MI.

For the AI, the sign and amplitude of the ac-Stark shift were constant in our experiments because the lattice beams were far detuned ($-|\Delta_\text{a}|$) from any atomic resonance (Figure \ref{fig:levels}a). Here, the ``$-$'' sign indicates that the lattice beams were red detuned from the nearest transition. On the other hand, for the MI, the amplitude and sign of the induced ac-Stark shift depended on the specific rovibronic, hyperfine and Zeeman state of the MI. Within the frequency range of the optical lattice used in the present experiments, a range of different spectroscopic transitions were accessible (Figure \ref{fig:levels}b) leading to different detunings ($\pm |\Delta_{m}|$) depending on the exact state of the MI.

Figure \ref{fig:cartoon}b shows a schematic of two different configurations of forces on the two ions used in our experiment. Since the AI was always red detuned, it always experienced a force in the direction of higher lattice intensity (high-field seeking configuration). For the MI, however, both high-field seeking and low-field seeking (blue detuning) scenarios were possible (upper and lower panels in Figure \ref{fig:cartoon}b). We set the relative phase of the forces by controlling the distance between the two ions, $d$. By placing the two ions at a separation that corresponded to an integer multiple of lattice nodes, $d=n\lambda/2$ ($n \in \mathbb{N}$ is an integer number and $\lambda$ is the wavelength of the lattice lasers), the ions experienced the same intensity gradient from the lattice. We denote this lattice configuration as "same phase" (SP). In the SP scenario, the forces were in the same (opposite) direction when the MI was red (blue) detuned (SP$_\textrm{R}$ and SP$_\textrm{B}$ in Figure \ref{fig:cartoon}b). By changing the two-ion distance by half a lattice spacing, $\Delta d=\pm\lambda/4$, the ions experienced an opposite gradient of the lattice field, i.e., the opposite phase (OP) of the lattice. In the OP lattice configuration, the forces were in the same (opposite) direction when the MI was blue (red) detuned (OP$_\textrm{R}$ and OP$_\textrm{B}$ in Figure \ref{fig:cartoon}b).

In the case when the forces interfered constructively (same directions), the resulting motional-excitation signal was stronger than in the case of destructive interference (opposite directions). This by itself did not yield information about the sign of the detuning since the ac-Stark shift varied greatly between different states such that a priori the strength of the signal could not be predicted. In order to disentangle the strength and sign of the ac-Stark shift, we compared the SP and OP configurations by varying the two-ion distance and thus obtained both the magnitude and sign of the ac-Stark shift. As the detuning ($\Delta_\text{m}$) changes across a resonance in the molecule, the direction of the induced ac-Stark shift (and hence the dipole force) changes while the direction of the shift (and hence the force) on the AI remains constant (see Figure \ref{fig:cartoon}c). As will be shown below, this information proves extremely valuable for distinguishing states in a complex energy level structure.

\subsection*{Experiment}

A detailed description of our experimental apparatus was given in Ref. \cite{meir19a, sinhal20}, a schematic representation is shown in Figure \ref{fig:cartoon}a and a brief overview is given in the Methods section.

\begin{figure}
	\centering
	\includegraphics[width=\linewidth,trim={0cm 0cm 0cm 0cm},clip]{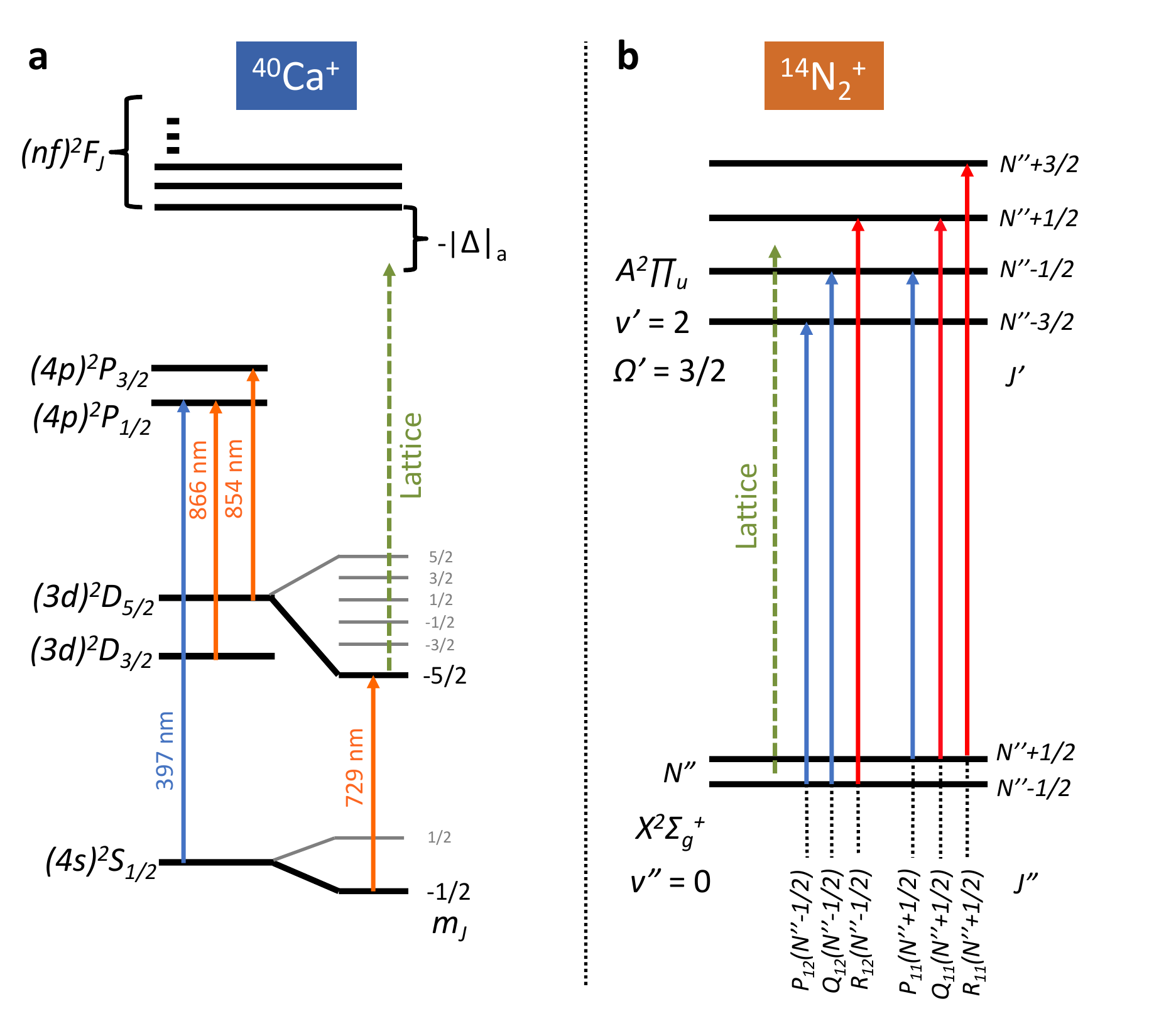}
 	\caption{Energy-level diagrams. a) Reduced energy-level diagram (not to scale) of $^{40}$Ca$^+$ relevant for Doppler and resolved-sideband cooling as well as motional state readout (solid arrows indicate transition frequencies). Motional excitation was initiated by a 1D lattice (dashed arrow indicates the laser frequency not to scale) which mainly interacted with the $(nf)F_{J}(m=-5/2)$ states (where $n\ge4$ and $J=7/2,5/2$) when Ca$^+$ was shelved in the $(3d)^2D_{5/2}(m=-5/2)$ level. The transition wavelength to the $F$ states is smaller than 184 nm such that the lattice at 789 nm is far red detuned. b) Simplified energy-level diagram (not to scale) of the $A^2\Pi_u (v' = 2) \leftarrow X^2\Sigma_g^+ (v'' = 0)$ band of $^{14}$N$_2^+$. Solid arrows indicate the possible transitions from a single rotational level (split into two spin-rotation components) of the ground vibronic state. Red (blue) arrows indicates red (blue) detuning of the lattice frequency (dashed green) with respect to the relevant transitions.}
	\label{fig:levels}
\end{figure}

The relevant energy levels of $^{40}$Ca$^+$ and $^{14}$N$_2^+$ are shown in Figure \ref{fig:levels}. With lattice beams linearly polarized and parallel to the magnetic-field direction, the lattice mainly interacted with high-lying $(nf)F_{7/2}(m=-5/2)$ and $(nf)F_{5/2}(m=-5/2)$ levels in Ca$^+$ since the AI was prepared in the metastable $(3d)~^2D_{5/2}(m=-5/2)$ state. Here, $m$ denotes the magnetic quantum number. Even tough the lattice was highly red-detuned from these AI levels ($|\Delta_{a}|>1250$ THz), their contribution to the ac-Stark shift was not negligible (Methods). For N$_2^+$, the lattice was nearly resonant with the $A^2\Pi_u (v' = 2) \leftarrow X^2\Sigma_g^+ (v'' = 0)$ vibronic transition where $v$ denotes the vibrational quantum number and $'$ ($''$) stand for the upper (lower) level of the transition. The detuning was dependent on the exact rotational, fine, hyperfine and Zeeman state of the MI. Figure \ref{fig:levels}b shows the detuning of the lattice on the P($J''$), Q($J''$), and R($J''$) lines initiating from the two spin-rotation components of the rovibronic ground state. Here, $J''$ is the quantum number of the total molecular angular momentum without nuclear spin of the MI in its $X^2\Sigma_g^+ (v'' = 0,N'',J''=N''\pm1/2)$ ground state and P,Q,R stand for spectroscopic branches with $\Delta J = J' - J'' = -1,0,1$. 

An ODF pulse was applied to detect the state of the molecule. In this pulse, the lattice lasers were turned on for a duration of 3 ms to excite motion in the crystal. The motional state of the IP mode was subsequently probed on the AI by Rabi sideband thermometry on a blue sideband of the  $D_{5/2}(m=-5/2)$ $\rightarrow$ $S_{1/2}(m=-1/2)$ transition \cite{meekhof96a,leibfried03a}. The frequency and contrast of the resulting Rabi-oscillation signal was proportional to the ODF strength (Methods).

In order to obtain both the sign and the magnitude of the ac-Stark shift experienced by the MI, the Rabi-oscillation experiment was performed twice. Once with an ion distance of $d=n\lambda/2$ where $n=19$ and $\lambda \approx 789$ nm corresponding to the SP configuration ($f_{IP}^{SP} = 695$ kHz) and a second time with the ion distance shifted by $\Delta d = +\lambda/4$ corresponding to the OP configuration ($f_{IP}^{OP} = 668$ kHz). The configurations were changed by adjusting the voltages on the ion-trap end-cap electrodes. The relative uncertainty in determining the ion-ion distance was estimated to be $\delta d / d = 10^{-3}$ (Methods). 

\begin{figure}
	\centering
	\includegraphics[width=\linewidth,trim={0cm 0cm 0cm 0cm},clip]{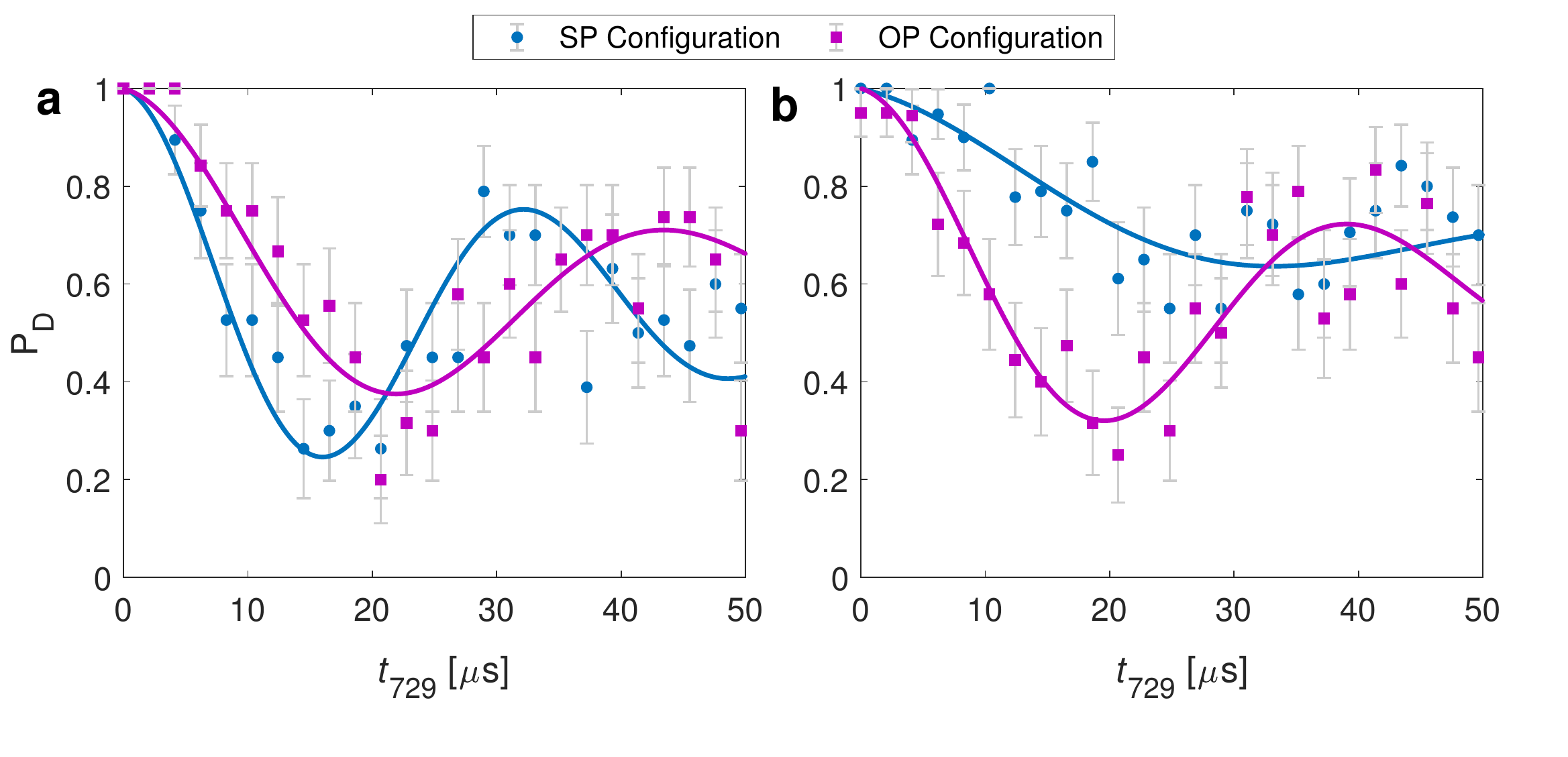}
 	\caption{Rabi thermometry. Rabi oscillation signals for the SP (blue) and OP (purple) lattice configurations after an ODF pulse for two different molecular states (panel (a) and panel (b)). Each data point is an average of 20 ODF excitations. Error bars represent $1\sigma$ binomial uncertainties. The solid curves represent fits to the data as described in the Methods section from which the ac-Stark shifts were extracted. a) The stronger SP signal (larger frequency and amplitude of the Rabi oscillation) suggest that the lattice laser ($\lambda = 789.71$ nm, in both experiments) was red detuned from the closest molecular transition. By comparing the measured amplitude and sign of the ac-Stark shift with theory (Methods), the molecular state could unambiguously be be identified as $N''=6$, $J'' = 11/2$. b) The stronger OP signal suggests that the lattice laser was blue detuned from the closest molecular transition. Therefore, we can exclude all rotational states $N''\le4$ (see Figure \ref{fig:partial}).}
	\label{fig:exsig}
\end{figure}

Example of Rabi-oscillation signals from two SP/OP experiments are shown in Figure \ref{fig:exsig}. The blue-sideband-pulse duration, $t_{729}$, was scanned and averaged 20 times for each data point to retrieve two Rabi-oscillation signals in each experiment, one for the SP (blue) and the other one for the OP (purple) configurations. The frequency and contrast of the Rabi signals were observed to be higher for the SP configuration compared to the OP configuration in Figure \ref{fig:exsig}a, implying that the ODF and overall motional excitation was stronger for the SP configuration. This indicated that the detuning of the lattice frequency from a molecular resonance was to the red in this experiment (see upper panel of Figure \ref{fig:cartoon}b). The resulting Rabi oscillations were fitted to a phenomenological function which was used to extract the amplitude of the ac-Stark shift (Methods). Based on the obtained values for the phase and amplitude, the molecular state could unambiguously be identified as $N''=6$, $J''=11/2$, as illustrated in the following section. 

In Figure \ref{fig:exsig}b, the opposite situation occurred for the same settings of the lattice laser beams. Here, the observed frequency and contrast of the Rabi signals were higher for the OP configuration. This implied that the lattice frequency was blue-detuned from the closest molecular resonance. Below, we show how based on this observation only, one could exclude all molecular states with $N''\le4$.

\subsection*{Results and discussion}

\begin{figure*}
	\centering
	\includegraphics[width=\linewidth,trim={0cm 0cm 0cm 0cm},clip]{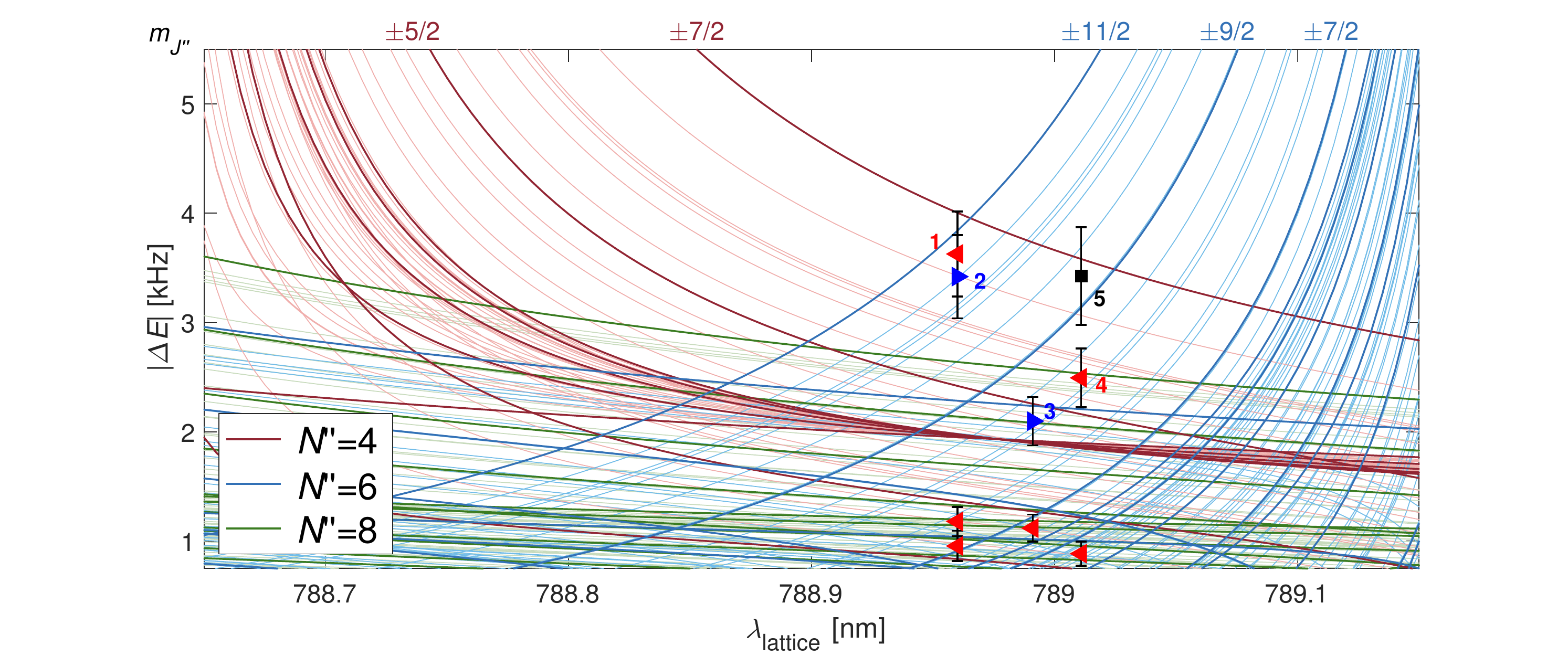}
 	\caption{Ac Stark shifts. Absolute magnitude of the ac-Stark shifts for molecules in the $N'' = 4,6,8$ rotational states (red, blue and green lines) at a lattice-laser wavelength around 789~nm. Thick lines correspond to the ac-Stark shift of Zeeman components of the $I=0$ nuclear-spin isomer without hyperfine structure (labeled by their magnetic quantum numbers $m_{J''}$ on top of the figure) while thin lines correspond to the hyperfine-Zeeman components of the $I=2$ isomer. Blue (red) triangles represent experiments in which the measured ac-Stark shifts were stronger in the OP (SP) configuration corresponding to a lattice blue (red) detuned from the transition. The black square represents an experiment without the phase information. Error bars are the combined $1 \sigma$ uncertainties of the fit to the data (Methods) and the uncertainty of the lattice laser power ($\sim 10 \%$). The ac-Stark shifts originating from the $N''=0,2$ rotational levels are lower than the ones for the $N''=8$ state in this wavelength range and were omitted for clarity.
 	}
	\label{fig:N4and6}
\end{figure*}

Our method applied to the identification of specific spin-rotation and Zeeman levels of N$_2^+$ molecules in the $X^2\Sigma_g^+ (v'' = 0)$ electronic and vibrational ground state is exemplified in Figure \ref{fig:N4and6} with the $N'' = 4$ and $N'' = 6$ states. By tuning the lattice-laser wavelength near 789 nm, the detection was sensitive to states originating from the $N'' = 4$, $J'' = 7/2$ spin-rotation manifold and induced a significant ac-Stark shift through the Q$_{12}(7/2)$ transition at 788.624 nm (red lines in Figure \ref{fig:N4and6}). For these states, the lattice frequency was red detuned with respect to the transition frequency. At the same wavelength, the detection was also sensitive to states originating from the $N'' = 6$, $J'' = 11/2$ spin-rotation manifold. These states induced a significant ac-Stark shift through the Q$_{12}(11/2)$ transition at 789.1872 nm (blue lines in Figure \ref{fig:N4and6}). For these states, the lattice frequency was blue detuned with respect to the transition frequency. 

In the present experiments, molecules in the electronic and vibrational ground state were primarily created in rotational states with $N''\le8$ by a rotationally unselective photoionization scheme. In each experiment, the molecules were randomly initialized in one of the 540 corresponding hyperfine Zeeman states. The detection protocol was then applied to extract the amplitude and phase of the ODF signal. In many experiments, the signal was very low, indicating that the state of the molecule was not within the subspace of states which could effectively be traced at the chosen lattice-laser wavelength (examples are shown by the red triangles in the bottom of Figure \ref{fig:N4and6}). However, for some experiments (labelled 1-3 in Figure \ref{fig:N4and6}), the ODF amplitude was large enough indicating that the molecule was either in the $N''=6$, $J''=11/2$ or the $N''=4$, $J''=7/2$ spin-rotational states. The added phase information enabled us to distinguish between these two spin-rotation manifolds (blue and red triangles respectively). The black square exemplifies an experiment in which the phase information was lacking and, therefore, an unambiguous identification of the molecular state was not possible. 

For the experiment marked ``1'' in Figure \ref{fig:N4and6}, the state was identified as $N''=4$, $J''=7/2$. The measured and the predicted ac-Stark shifts agree within 1$\sigma$ with only 6 hyperfine Zeeman states (see Table \ref{tab:states}). Thus, we can exclude 534 out of 540 possible states for a molecule with $N''\le8$. Even if we assume a discrepancy of $2\sigma$ between theory and our measured ac-Stark shifts, we can exclude 528 of the 540 possible states of the molecule (Table \ref{tab:states}). For the experiment marked ``2'' in Figure \ref{fig:N4and6}, the state was identified as $N''=6$, $J''=11/2$. Here again, only 6 (12) hyperfine Zeeman states agree with theory within 1$\sigma$ (2$\sigma$) thus excluding 99\% (98\%) of the possible states (Table \ref{tab:states}). 

The experiments marked ``3'' and ``4'' in Figure \ref{fig:N4and6} also show opposite phases at a comparable amplitude. While in experiment ``3'' we identified the state to be one of 4 (6) hyperfine Zeeman states of the $N''=6$, $J''=11/2$ spin-rotation manifold, experiment ``4'' is associated with 22 (42) hyperfine Zeeman states of the $N''=4$, $J''=7/2$ and $N''=8$, $J''=17/2$ spin-rotation manifolds (Table \ref{tab:states}). These experiments exemplified that using the phase information, it is possible to exclude more than 50\% (here $\sim$85\%) of the states compared to when only the amplitude information is available. 

\begin{table}
\begin{tabular}{| l |c |c |r |r |r |r |} \hline
Experiment No. & Agreement & $N''$ & $J''$ & $I''$ & $F''$ & $m_{J''/F''}$  \\ \hline
\textbf{1}  & $< 1 \sigma$     & \textbf{4} &  \textbf{7/2}  & 0 &   & $\pm 7/2$\\
   &                &   &  & 2 & 11/2  & $\pm 11/2$\\
   &                &   &  & 2 & 9/2   & $\pm 9/2$\\
   & $< 2 \sigma$     & \textbf{} &  & 2 & 11/2  & $\pm 9/2$\\
   &                &   &  & 2 & 7/2   & $\pm 7/2$\\
   &                &   &  & 2 & 3/2   & $\pm 3/2$\\ \hline
\textbf{2}  & $< 1 \sigma$     & \textbf{6} & \textbf{11/2}  & 2 & 13/2  & $\pm 13/2$ \\
   &                &   &  & 2 & 11/2  & $\pm 11/2$ \\
   &                &   &  & 2 & 7/2  & $\pm 7/2$ \\
   & $< 2 \sigma$     &  &  & 0 &  & $\pm 11/2$ \\
   &                &   & & 2 & 15/2  & $\pm 15/2$ \\
   &                &   &  & 2 & 9/2   & $\pm 9/2$ \\ \hline
\textbf{3}  & $< 1 \sigma$     & \textbf{6} & \textbf{11/2}  & 2 & 15/2  & $\pm 11/2$\\
   &                &   & & 2 & 9/2   & $\pm 7/2$\\
   & $< 2 \sigma$     &  &  & 2 & 11/2  & $\pm 9/2$\\ \hline
\textbf{4}  & $< 1 \sigma$     & \textbf{4} &  \textbf{7/2} & 2 & 11/2   & $\pm 9/2$\\
   &                &   & & 2 & 9/2   & $\pm 7/2$\\
   &                &   & & 2 & 7/2   & $\pm 7/2$\\
   &                &   & & 2 & 5/2   & $\pm 5/2$\\
   &                &   & & 2 & 3/2   & $\pm 3/2$\\ 
   &                & \textbf{8} & \textbf{17/2} & 0 &  & $\pm 17/2$\\
   &                &   & & 2 & 21/2  & $\pm 21/2$\\
   &                &   & & 2 & 19/2  & $\pm 19/2$\\
   &                &   & & 2 & 17/2  & $\pm 17/2$\\
   &                &   & & 2 & 15/2  & $\pm 15/2$ \\
   &                &   & & 0 & 13/2  & $\pm 13/2$ \\ \hline
\end{tabular}
\caption{Molecular state identification. The numbers in the first column correspond to the labels shown in Figure \ref{fig:N4and6}. The second column indicates the level of agreement between experiment and theory for the given state (For experiment 4, only states within 1$\sigma$ agreement are given. An additional 20$\times$2 states within 2$\sigma$ agreement are omitted). $N''$, $J''$, $I''$, $F''$ and $m_{J''/F''}$ stand for the rotational, spin-rotational, nuclear-spin, hyperfine and magnetic quantum numbers of N$_2^+$ in the electronic and vibrational ground state.}
\label{tab:states}
\end{table}

\begin{figure}
	\centering
	\includegraphics[width=\linewidth,trim={0cm 0cm 0cm 0cm},clip]{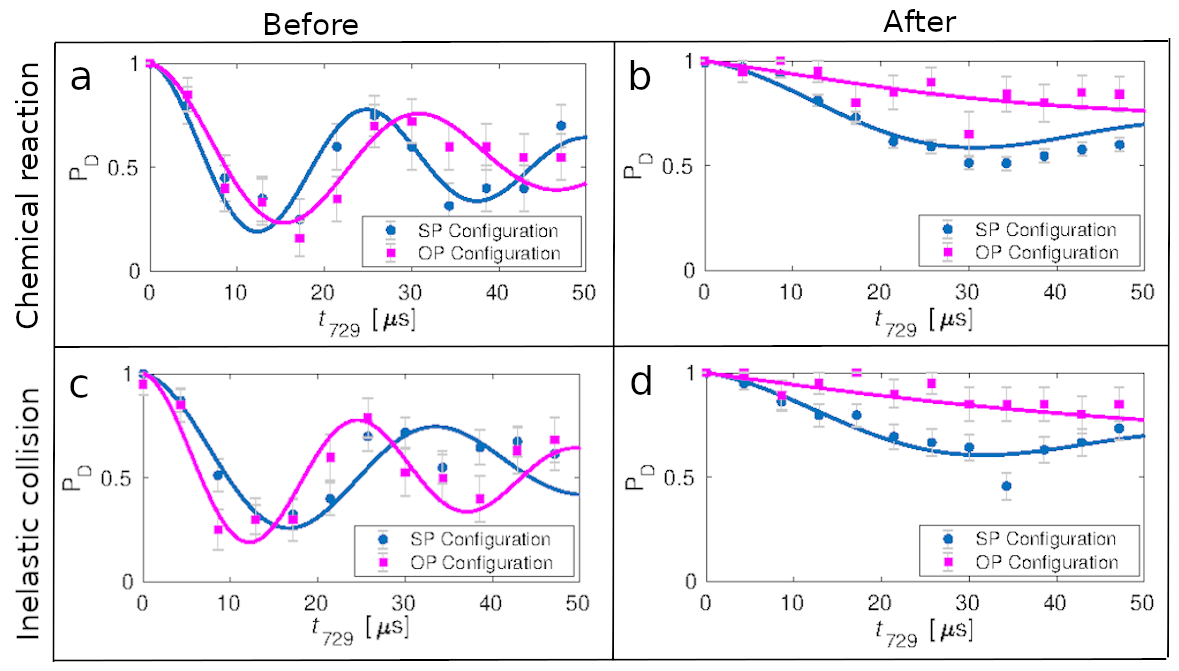}
 	\caption{Tracing molecular states during inelastic processes. State dynamics of two molecules (experiment No. ``1'' and ``2'' in Figure \ref{fig:N4and6}) probed by the present SP/OP protocol under a chemical reaction (a $\rightarrow$ b) and a quantum jump (c $\rightarrow$ d). a $\rightarrow$ b): An N$_2^+$ molecule in the $N'' = 4, J'' = 7/2$ state reacts with a background-gas H$_2$ molecule and converts into N$_2$H$^+$. c $\rightarrow$ d) An N$_2^+$ molecule in $N'' = 6$, $J'' = 11/2$ changes its state (likely to $N''=4$) due to an inelastic collision or photon scattering.}
	\label{fig:N4sig}
\end{figure}

To further illustrate the capabilities of the present state-detection protocol, a state-sensitive chemical reaction experiment was implemented to observe how N$_2^+$ molecules in a determined state react with H$_2$ background gas molecules to form N$_2$H$^+$. Figure \ref{fig:N4sig} shows the results of such an experiment. An N$_2^+$ molecule was identified to be in the $N''=4$, $J''=7/2$ spin-rotational state (panel a). The molecule eventually reacted with background gas and turned into N$_2$H$^+$. The change of the Rabi-oscillation signal from panel a to panel b indicated the change of the chemical composition of the molecule which was verified by mass spectrometry (Methods). Due to the mass change of the crystal in the reaction, the frequency difference of the lattice beams, $\Delta f$, had to be adjusted to match the new crystal frequency $f_{IP}^{SP}(N_2H^+) = 690$ kHz (Methods). The energy-level structure of N$_2$H$^+$ is not sufficiently known to be able to clearly identify the molecular quantum state generated in the reaction. However, the phase information indicated that the lattice beams were red detuned from the relevant resonance in the N$_2$H$^+$ molecule. This experiment exemplifies the basic possibility to perform state-to-state chemical reaction experiments involving polyatomic species.

Following the same principle, inelastic, i.e., state-changing processes of a single molecule can also be traced. In panel c and d of Figure \ref{fig:N4sig}, the results of such an experiment are shown. An N$_2^+$ ion in the $N'' = 6$,  $J'' = 11/2$ spin-rotational state underwent a quantum-jump to a different rotational state as can be seen by the change in the amplitude and phase of the Rabi-oscillation signal. The change of state could have been caused by either an inelastic collision with a background-gas molecule or by the scattering of a photon from the lattice laser \cite{sinhal20}. Prior to the quantum jump, the OP configuration showed a stronger signal than the SP configuration while after the collision, the OP configuration showed a weaker signal than the SP configuration. This suggests that the molecule underwent a rotational-state change, most probably to the $N''=4$ state though the low amplitude of the signal does not allow us to unambiguously exclude other rotational states. The state of the molecule after the quantum jump could have been tested with higher fidelity by changing the lattice laser frequency to a value at which the assumed state would result in an increased signal amplitude.

The experiments presented so far relied on prior knowledge of the strengths and positions of molecular transitions to positively identify specific states of a molecule by comparing the experimentally obtained ac-Stark shift to theory. In a variety of molecules, this information will not be completely available. For the common situation in which only the frequencies of the transitions are known, e.g., from a prediction based on known spectroscopic constants, we propose and demonstrate an adaption of the present method for ``partial state readout'' \cite{patterson18} that can be used to exclude a large subset of molecular states simultaneously and provide a non-destructive spectroscopic signal. 

\begin{figure}
	\centering
	\includegraphics[width=\linewidth,trim={0cm 0cm 0cm 0cm},clip]{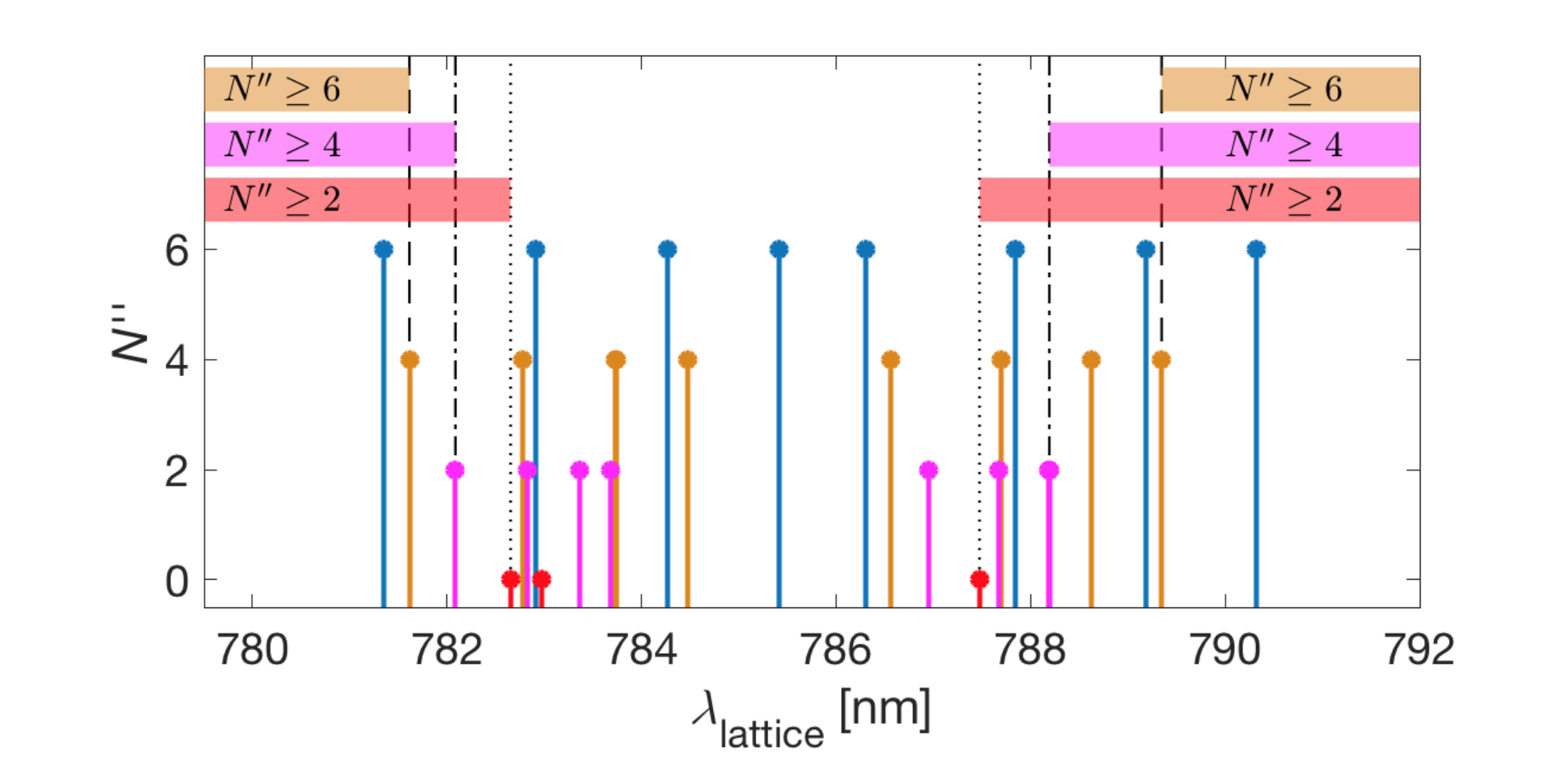}
 	\caption{State identification within a dense level structure. Stick spectrum showing the positions of the allowed P, Q, and R transitions of the $A^2\Pi_u (v' = 2) \leftarrow X^2\Sigma_g^+ (v'' = 0)$ band of N$_2^+$originating from the rotational levels $N''$ = 0, 2, 4 and 6. A subset of these states can be excluded by determining the effective detuning at specific lattice wavelengths using an SP/OP experiment. The dotted, dash-dotted and dashed lines show the lattice wavelengths that enable the exclusion of rotational states $N'' = 0$, $N'' \le 2$ and $N'' \leq 4$ respectively. The detuning can be chosen either blue of the transitions (782.6, 782.1 and 781.6 nm to exclude $N'' = 0$, $N'' \leq 2$ and $N'' \leq 4$ respectively) or red (787.5, 788.2, and 789.4 nm to exclude $N'' = 0$, $N'' \leq 2$ and $N'' \leq 4$). By choosing the detuning opposite to the detuning of an excited state of interest, the initial and final states are easily distinguishable in an SP/OP test with no available information about the hyperfine or Zeeman structure of either upper or lower states or the specific transition strengths.}
	\label{fig:partial}
\end{figure}

An example is given for the present case of N$_2^+$. The positions of all spectral lines belonging to the $A^2\Pi_u (v' = 2) \leftarrow X^2\Sigma_g^+ (v'' = 0)$ transitions up to $N'' = 6$ colour coded by rotational state are shown in Figure \ref{fig:partial}. As can be seen in the figure, a lattice-laser wavelength larger than 789.4 nm (dashed line) is red-detuned with respect to all transitions with $N'' \leq 4$. 
Therefore, detecting a blue detuning in an SP/OP experiment excludes this entire manifold regardless of its substructure or strengths.
In Figure \ref{fig:exsig}b, we show a demonstration for this ``partial state readout''. We used a lattice wavelength of 789.71 nm and detected a stronger OP than SP signal which implied that the lattice was blue detuned to the molecular transition thus allowing us to exclude all states with rotational quantum number $N''\le4$ for this molecule.

Conversely, the present protocol can be used as a non-destructive readout for spectroscopic excitations when exciting from a partially known initial state, e.g., $N'' \leq 4$ ($v'' = 0$) to a long-lived excited state with a known detuning with respect to the lattice such as the $(v' = 1)$ state of the $X^2\Sigma_g^+$ ~ ground state. At 789.7 nm, the lattice laser is blue-detuned from the closest transitions in ($v' = 1$) and one can detect a successful spectroscopic excitation from any states of $N'' \leq 4$ as a change in the detuning (i.e. the sign of the ac-Stark shift) even with no available information about the hyperfine or Zeeman structure of either upper or lower states or the transition strengths. Such an approach will greatly enhance the possibilities for non-destructive spectroscopic experiments in complex molecules.

\subsection*{Summary}
To conclude, we have demonstrated an experimental protocol for molecular-state detection relying on phase-sensitive optical-dipole forces. The present approach incorporates both the amplitude and the phase of the detection signal in the molecular-state identification. Using this protocol, Zeeman levels of rotational states of N$_2^+$ were identified in a region where information about only the amplitude of the signal leads to ambiguity in the state identification. The present method was also demonstrated to trace reactive and state-changing collisions of single molecules with partial state selectively.  

We also showed how the phase information provided by the present method can be used for a partial state determination in situations in which only incomplete spectroscopic information is available on the molecule. The present work thus introduces important new tools towards non-destructive state identification and spectroscopy of complex molecular systems.

\subsection*{Acknowledgements}
We thank M. S. Safronova for providing accurate values for the Ca$^+$ polarizability and P. Stra\v{n}\'{a}k for calculating the polarizability of N$_2^{2+}$. This work has been supported by the Swiss National Science Foundation as part of the National Centre of Competence in Research, Quantum Science and Technology (NCCR-QSIT), grant nr. CRSII5\_183579, and by the University of Basel.

\subsection*{Author contributions}
KN and ZM performed the experiments and the analysis of the data. MS developed parts of the theory underlying the analysis. SW conceived and supervised the project. All authors contributed to writing the manuscript.


\begin{thebibliography}{44}%
\makeatletter
\providecommand \@ifxundefined [1]{%
 \@ifx{#1\undefined}
}%
\providecommand \@ifnum [1]{%
 \ifnum #1\expandafter \@firstoftwo
 \else \expandafter \@secondoftwo
 \fi
}%
\providecommand \@ifx [1]{%
 \ifx #1\expandafter \@firstoftwo
 \else \expandafter \@secondoftwo
 \fi
}%
\providecommand \natexlab [1]{#1}%
\providecommand \enquote  [1]{``#1''}%
\providecommand \bibnamefont  [1]{#1}%
\providecommand \bibfnamefont [1]{#1}%
\providecommand \citenamefont [1]{#1}%
\providecommand \href@noop [0]{\@secondoftwo}%
\providecommand \href [0]{\begingroup \@sanitize@url \@href}%
\providecommand \@href[1]{\@@startlink{#1}\@@href}%
\providecommand \@@href[1]{\endgroup#1\@@endlink}%
\providecommand \@sanitize@url [0]{\catcode `\\12\catcode `\$12\catcode
  `\&12\catcode `\#12\catcode `\^12\catcode `\_12\catcode `\%12\relax}%
\providecommand \@@startlink[1]{}%
\providecommand \@@endlink[0]{}%
\providecommand \url  [0]{\begingroup\@sanitize@url \@url }%
\providecommand \@url [1]{\endgroup\@href {#1}{\urlprefix }}%
\providecommand \urlprefix  [0]{URL }%
\providecommand \Eprint [0]{\href }%
\providecommand \doibase [0]{http://dx.doi.org/}%
\providecommand \selectlanguage [0]{\@gobble}%
\providecommand \bibinfo  [0]{\@secondoftwo}%
\providecommand \bibfield  [0]{\@secondoftwo}%
\providecommand \translation [1]{[#1]}%
\providecommand \BibitemOpen [0]{}%
\providecommand \bibitemStop [0]{}%
\providecommand \bibitemNoStop [0]{.\EOS\space}%
\providecommand \EOS [0]{\spacefactor3000\relax}%
\providecommand \BibitemShut  [1]{\csname bibitem#1\endcsname}%
\let\auto@bib@innerbib\@empty
\bibitem [{\citenamefont {Wright}\ \emph {et~al.}(2019)\citenamefont {Wright},
  \citenamefont {Beck}, \citenamefont {Debnath}, \citenamefont {Amini},
  \citenamefont {Nam}, \citenamefont {Grzesiak}, \citenamefont {Chen},
  \citenamefont {Pisenti}, \citenamefont {Chmielewski}, \citenamefont {Collins}
  \emph {et~al.}}]{wright19}%
  \BibitemOpen
  \bibfield  {author} {\bibinfo {author} {\bibfnamefont {K.}~\bibnamefont
  {Wright}}, \bibinfo {author} {\bibfnamefont {K.}~\bibnamefont {Beck}},
  \bibinfo {author} {\bibfnamefont {S.}~\bibnamefont {Debnath}}, \bibinfo
  {author} {\bibfnamefont {J.}~\bibnamefont {Amini}}, \bibinfo {author}
  {\bibfnamefont {Y.}~\bibnamefont {Nam}}, \bibinfo {author} {\bibfnamefont
  {N.}~\bibnamefont {Grzesiak}}, \bibinfo {author} {\bibfnamefont {J.-S.}\
  \bibnamefont {Chen}}, \bibinfo {author} {\bibfnamefont {N.}~\bibnamefont
  {Pisenti}}, \bibinfo {author} {\bibfnamefont {M.}~\bibnamefont
  {Chmielewski}}, \bibinfo {author} {\bibfnamefont {C.}~\bibnamefont
  {Collins}},  \emph {et~al.},\ }\href@noop {} {\bibfield  {journal} {\bibinfo
  {journal} {Nat. Commun.}\ }\textbf {\bibinfo {volume} {10}},\ \bibinfo
  {pages} {1} (\bibinfo {year} {2019})}\BibitemShut {NoStop}%
\bibitem [{\citenamefont {Friis}\ \emph {et~al.}(2018)\citenamefont {Friis},
  \citenamefont {Marty}, \citenamefont {Maier}, \citenamefont {Hempel},
  \citenamefont {Holz{\"a}pfel}, \citenamefont {Jurcevic}, \citenamefont
  {Plenio}, \citenamefont {Huber}, \citenamefont {Roos}, \citenamefont {Blatt}
  \emph {et~al.}}]{friis18}%
  \BibitemOpen
  \bibfield  {author} {\bibinfo {author} {\bibfnamefont {N.}~\bibnamefont
  {Friis}}, \bibinfo {author} {\bibfnamefont {O.}~\bibnamefont {Marty}},
  \bibinfo {author} {\bibfnamefont {C.}~\bibnamefont {Maier}}, \bibinfo
  {author} {\bibfnamefont {C.}~\bibnamefont {Hempel}}, \bibinfo {author}
  {\bibfnamefont {M.}~\bibnamefont {Holz{\"a}pfel}}, \bibinfo {author}
  {\bibfnamefont {P.}~\bibnamefont {Jurcevic}}, \bibinfo {author}
  {\bibfnamefont {M.~B.}\ \bibnamefont {Plenio}}, \bibinfo {author}
  {\bibfnamefont {M.}~\bibnamefont {Huber}}, \bibinfo {author} {\bibfnamefont
  {C.}~\bibnamefont {Roos}}, \bibinfo {author} {\bibfnamefont {R.}~\bibnamefont
  {Blatt}},  \emph {et~al.},\ }\href@noop {} {\bibfield  {journal} {\bibinfo
  {journal} {Phys. Rev. X}\ }\textbf {\bibinfo {volume} {8}},\ \bibinfo {pages}
  {021012} (\bibinfo {year} {2018})}\BibitemShut {NoStop}%
\bibitem [{\citenamefont {Arute}\ \emph {et~al.}(2019)\citenamefont {Arute},
  \citenamefont {Arya}, \citenamefont {Babbush}, \citenamefont {Bacon},
  \citenamefont {Bardin}, \citenamefont {Barends}, \citenamefont {Biswas},
  \citenamefont {Boixo}, \citenamefont {Brandao}, \citenamefont {Buell} \emph
  {et~al.}}]{arute19}%
  \BibitemOpen
  \bibfield  {author} {\bibinfo {author} {\bibfnamefont {F.}~\bibnamefont
  {Arute}}, \bibinfo {author} {\bibfnamefont {K.}~\bibnamefont {Arya}},
  \bibinfo {author} {\bibfnamefont {R.}~\bibnamefont {Babbush}}, \bibinfo
  {author} {\bibfnamefont {D.}~\bibnamefont {Bacon}}, \bibinfo {author}
  {\bibfnamefont {J.~C.}\ \bibnamefont {Bardin}}, \bibinfo {author}
  {\bibfnamefont {R.}~\bibnamefont {Barends}}, \bibinfo {author} {\bibfnamefont
  {R.}~\bibnamefont {Biswas}}, \bibinfo {author} {\bibfnamefont
  {S.}~\bibnamefont {Boixo}}, \bibinfo {author} {\bibfnamefont {F.~G.}\
  \bibnamefont {Brandao}}, \bibinfo {author} {\bibfnamefont {D.~A.}\
  \bibnamefont {Buell}},  \emph {et~al.},\ }\href@noop {} {\bibfield  {journal}
  {\bibinfo  {journal} {Nature}\ }\textbf {\bibinfo {volume} {574}},\ \bibinfo
  {pages} {505} (\bibinfo {year} {2019})}\BibitemShut {NoStop}%
\bibitem [{\citenamefont {Brewer}\ \emph {et~al.}(2019)\citenamefont {Brewer},
  \citenamefont {Chen}, \citenamefont {Hankin}, \citenamefont {Clements},
  \citenamefont {Chou}, \citenamefont {Wineland}, \citenamefont {Hume},\ and\
  \citenamefont {Leibrandt}}]{brewer19}%
  \BibitemOpen
  \bibfield  {author} {\bibinfo {author} {\bibfnamefont {S.~M.}\ \bibnamefont
  {Brewer}}, \bibinfo {author} {\bibfnamefont {J.-S.}\ \bibnamefont {Chen}},
  \bibinfo {author} {\bibfnamefont {A.~M.}\ \bibnamefont {Hankin}}, \bibinfo
  {author} {\bibfnamefont {E.~R.}\ \bibnamefont {Clements}}, \bibinfo {author}
  {\bibfnamefont {C.~W.}\ \bibnamefont {Chou}}, \bibinfo {author}
  {\bibfnamefont {D.~J.}\ \bibnamefont {Wineland}}, \bibinfo {author}
  {\bibfnamefont {D.~B.}\ \bibnamefont {Hume}}, \ and\ \bibinfo {author}
  {\bibfnamefont {D.~R.}\ \bibnamefont {Leibrandt}},\ }\href@noop {} {\bibfield
   {journal} {\bibinfo  {journal} {Phys. Rev. Lett.}\ }\textbf {\bibinfo
  {volume} {123}},\ \bibinfo {pages} {033201} (\bibinfo {year}
  {2019})}\BibitemShut {NoStop}%
\bibitem [{\citenamefont {Huntemann}\ \emph {et~al.}(2016)\citenamefont
  {Huntemann}, \citenamefont {Sanner}, \citenamefont {Lipphardt}, \citenamefont
  {\mbox{Chr.} Tamm},\ and\ \citenamefont {Peik}}]{huntemann16a}%
  \BibitemOpen
  \bibfield  {author} {\bibinfo {author} {\bibfnamefont {N.}~\bibnamefont
  {Huntemann}}, \bibinfo {author} {\bibfnamefont {C.}~\bibnamefont {Sanner}},
  \bibinfo {author} {\bibfnamefont {B.}~\bibnamefont {Lipphardt}}, \bibinfo
  {author} {\bibnamefont {\mbox{Chr.} Tamm}}, \ and\ \bibinfo {author}
  {\bibfnamefont {E.}~\bibnamefont {Peik}},\ }\href@noop {} {\bibfield
  {journal} {\bibinfo  {journal} {{Phys. Rev. Lett.}}\ }\textbf {\bibinfo
  {volume} {116}},\ \bibinfo {pages} {063001} (\bibinfo {year}
  {2016})}\BibitemShut {NoStop}%
\bibitem [{\citenamefont {Safronova}\ \emph {et~al.}(2018)\citenamefont
  {Safronova}, \citenamefont {Budker}, \citenamefont {DeMille}, \citenamefont
  {Kimball}, \citenamefont {Derevianko},\ and\ \citenamefont
  {Clark}}]{safronova18a}%
  \BibitemOpen
  \bibfield  {author} {\bibinfo {author} {\bibfnamefont {M.~S.}\ \bibnamefont
  {Safronova}}, \bibinfo {author} {\bibfnamefont {D.}~\bibnamefont {Budker}},
  \bibinfo {author} {\bibfnamefont {D.}~\bibnamefont {DeMille}}, \bibinfo
  {author} {\bibfnamefont {D.~F.~J.}\ \bibnamefont {Kimball}}, \bibinfo
  {author} {\bibfnamefont {A.}~\bibnamefont {Derevianko}}, \ and\ \bibinfo
  {author} {\bibfnamefont {C.~W.}\ \bibnamefont {Clark}},\ }\href@noop {}
  {\bibfield  {journal} {\bibinfo  {journal} {Rev. Mod. Phys.}\ }\textbf
  {\bibinfo {volume} {90}},\ \bibinfo {pages} {025008} (\bibinfo {year}
  {2018})}\BibitemShut {NoStop}%
\bibitem [{\citenamefont {DeMille}\ \emph {et~al.}(2017)\citenamefont
  {DeMille}, \citenamefont {Doyle},\ and\ \citenamefont
  {Sushkov}}]{demille17a}%
  \BibitemOpen
  \bibfield  {author} {\bibinfo {author} {\bibfnamefont {D.}~\bibnamefont
  {DeMille}}, \bibinfo {author} {\bibfnamefont {J.~M.}\ \bibnamefont {Doyle}},
  \ and\ \bibinfo {author} {\bibfnamefont {A.~O.}\ \bibnamefont {Sushkov}},\
  }\href@noop {} {\bibfield  {journal} {\bibinfo  {journal} {Science}\ }\textbf
  {\bibinfo {volume} {357}},\ \bibinfo {pages} {990} (\bibinfo {year}
  {2017})}\BibitemShut {NoStop}%
\bibitem [{\citenamefont {Liu}\ \emph {et~al.}(2018)\citenamefont {Liu},
  \citenamefont {Hood}, \citenamefont {Yu}, \citenamefont {Zhang},
  \citenamefont {Hutzler}, \citenamefont {Rosenband},\ and\ \citenamefont
  {Ni}}]{liu18}%
  \BibitemOpen
  \bibfield  {author} {\bibinfo {author} {\bibfnamefont {L.}~\bibnamefont
  {Liu}}, \bibinfo {author} {\bibfnamefont {J.}~\bibnamefont {Hood}}, \bibinfo
  {author} {\bibfnamefont {Y.}~\bibnamefont {Yu}}, \bibinfo {author}
  {\bibfnamefont {J.}~\bibnamefont {Zhang}}, \bibinfo {author} {\bibfnamefont
  {N.}~\bibnamefont {Hutzler}}, \bibinfo {author} {\bibfnamefont
  {T.}~\bibnamefont {Rosenband}}, \ and\ \bibinfo {author} {\bibfnamefont
  {K.-K.}\ \bibnamefont {Ni}},\ }\href@noop {} {\bibfield  {journal} {\bibinfo
  {journal} {Science}\ }\textbf {\bibinfo {volume} {360}},\ \bibinfo {pages}
  {900} (\bibinfo {year} {2018})}\BibitemShut {NoStop}%
\bibitem [{\citenamefont {Sikorsky}\ \emph {et~al.}(2018)\citenamefont
  {Sikorsky}, \citenamefont {Meir}, \citenamefont {Ben-shlomi}, \citenamefont
  {Akerman},\ and\ \citenamefont {Ozeri}}]{sikorsky18a}%
  \BibitemOpen
  \bibfield  {author} {\bibinfo {author} {\bibfnamefont {T.}~\bibnamefont
  {Sikorsky}}, \bibinfo {author} {\bibfnamefont {Z.}~\bibnamefont {Meir}},
  \bibinfo {author} {\bibfnamefont {R.}~\bibnamefont {Ben-shlomi}}, \bibinfo
  {author} {\bibfnamefont {N.}~\bibnamefont {Akerman}}, \ and\ \bibinfo
  {author} {\bibfnamefont {R.}~\bibnamefont {Ozeri}},\ }\href@noop {}
  {\bibfield  {journal} {\bibinfo  {journal} {Nat. Commun.}\ }\textbf {\bibinfo
  {volume} {9}},\ \bibinfo {pages} {920} (\bibinfo {year} {2018})}\BibitemShut
  {NoStop}%
\bibitem [{\citenamefont {D{\"o}rfler}\ \emph {et~al.}(2019)\citenamefont
  {D{\"o}rfler}, \citenamefont {Eberle}, \citenamefont {Koner}, \citenamefont
  {Tomza}, \citenamefont {Meuwly},\ and\ \citenamefont
  {Willitsch}}]{dorfler19b}%
  \BibitemOpen
  \bibfield  {author} {\bibinfo {author} {\bibfnamefont {A.~D.}\ \bibnamefont
  {D{\"o}rfler}}, \bibinfo {author} {\bibfnamefont {P.}~\bibnamefont {Eberle}},
  \bibinfo {author} {\bibfnamefont {D.}~\bibnamefont {Koner}}, \bibinfo
  {author} {\bibfnamefont {M.}~\bibnamefont {Tomza}}, \bibinfo {author}
  {\bibfnamefont {M.}~\bibnamefont {Meuwly}}, \ and\ \bibinfo {author}
  {\bibfnamefont {S.}~\bibnamefont {Willitsch}},\ }\href@noop {} {\bibfield
  {journal} {\bibinfo  {journal} {Nat. Commun.}\ }\textbf {\bibinfo {volume}
  {10}},\ \bibinfo {pages} {5429} (\bibinfo {year} {2019})}\BibitemShut
  {NoStop}%
\bibitem [{\citenamefont {Kimble}(2008)}]{kimble08a}%
  \BibitemOpen
  \bibfield  {author} {\bibinfo {author} {\bibfnamefont {H.~J.}\ \bibnamefont
  {Kimble}},\ }\href@noop {} {\bibfield  {journal} {\bibinfo  {journal}
  {Nature}\ }\textbf {\bibinfo {volume} {453}},\ \bibinfo {pages} {1023}
  (\bibinfo {year} {2008})}\BibitemShut {NoStop}%
\bibitem [{\citenamefont {Northup}\ and\ \citenamefont
  {Blatt}(2014)}]{northup14}%
  \BibitemOpen
  \bibfield  {author} {\bibinfo {author} {\bibfnamefont {T.}~\bibnamefont
  {Northup}}\ and\ \bibinfo {author} {\bibfnamefont {R.}~\bibnamefont
  {Blatt}},\ }\href@noop {} {\bibfield  {journal} {\bibinfo  {journal} {Nat.
  Photon.}\ }\textbf {\bibinfo {volume} {8}},\ \bibinfo {pages} {356} (\bibinfo
  {year} {2014})}\BibitemShut {NoStop}%
\bibitem [{\citenamefont {Wolf}\ \emph {et~al.}(2016)\citenamefont {Wolf},
  \citenamefont {Wan}, \citenamefont {Heip}, \citenamefont {Gebert},
  \citenamefont {Shi},\ and\ \citenamefont {Schmidt}}]{wolf16a}%
  \BibitemOpen
  \bibfield  {author} {\bibinfo {author} {\bibfnamefont {F.}~\bibnamefont
  {Wolf}}, \bibinfo {author} {\bibfnamefont {Y.}~\bibnamefont {Wan}}, \bibinfo
  {author} {\bibfnamefont {J.~C.}\ \bibnamefont {Heip}}, \bibinfo {author}
  {\bibfnamefont {F.}~\bibnamefont {Gebert}}, \bibinfo {author} {\bibfnamefont
  {C.}~\bibnamefont {Shi}}, \ and\ \bibinfo {author} {\bibfnamefont {P.~O.}\
  \bibnamefont {Schmidt}},\ }\href@noop {} {\bibfield  {journal} {\bibinfo
  {journal} {Nature}\ }\textbf {\bibinfo {volume} {530}},\ \bibinfo {pages}
  {457} (\bibinfo {year} {2016})}\BibitemShut {NoStop}%
\bibitem [{\citenamefont {Chou}\ \emph {et~al.}(2017)\citenamefont {Chou},
  \citenamefont {Kurz}, \citenamefont {Hume}, \citenamefont {Plessow},
  \citenamefont {Leibrandt},\ and\ \citenamefont {Leibfried}}]{chou17a}%
  \BibitemOpen
  \bibfield  {author} {\bibinfo {author} {\bibfnamefont {C.~W.}\ \bibnamefont
  {Chou}}, \bibinfo {author} {\bibfnamefont {C.}~\bibnamefont {Kurz}}, \bibinfo
  {author} {\bibfnamefont {D.~B.}\ \bibnamefont {Hume}}, \bibinfo {author}
  {\bibfnamefont {P.~N.}\ \bibnamefont {Plessow}}, \bibinfo {author}
  {\bibfnamefont {D.~R.}\ \bibnamefont {Leibrandt}}, \ and\ \bibinfo {author}
  {\bibfnamefont {D.}~\bibnamefont {Leibfried}},\ }\href@noop {} {\bibfield
  {journal} {\bibinfo  {journal} {Nature}\ }\textbf {\bibinfo {volume} {545}},\
  \bibinfo {pages} {203} (\bibinfo {year} {2017})}\BibitemShut {NoStop}%
\bibitem [{\citenamefont {Sinhal}\ \emph {et~al.}(2020)\citenamefont {Sinhal},
  \citenamefont {Meir}, \citenamefont {Najafian}, \citenamefont {Hegi},\ and\
  \citenamefont {Willitsch}}]{sinhal20}%
  \BibitemOpen
  \bibfield  {author} {\bibinfo {author} {\bibfnamefont {M.}~\bibnamefont
  {Sinhal}}, \bibinfo {author} {\bibfnamefont {Z.}~\bibnamefont {Meir}},
  \bibinfo {author} {\bibfnamefont {K.}~\bibnamefont {Najafian}}, \bibinfo
  {author} {\bibfnamefont {G.}~\bibnamefont {Hegi}}, \ and\ \bibinfo {author}
  {\bibfnamefont {S.}~\bibnamefont {Willitsch}},\ }\href@noop {} {\bibfield
  {journal} {\bibinfo  {journal} {Science}\ }\textbf {\bibinfo {volume}
  {367}},\ \bibinfo {pages} {1213} (\bibinfo {year} {2020})}\BibitemShut
  {NoStop}%
\bibitem [{\citenamefont {Chou}\ \emph {et~al.}(2020)\citenamefont {Chou},
  \citenamefont {Collopy}, \citenamefont {Kurz}, \citenamefont {Lin},
  \citenamefont {Harding}, \citenamefont {Plessow}, \citenamefont {Fortier},
  \citenamefont {Diddams}, \citenamefont {Leibfried},\ and\ \citenamefont
  {Leibrandt}}]{chou20a}%
  \BibitemOpen
  \bibfield  {author} {\bibinfo {author} {\bibfnamefont {C.}~\bibnamefont
  {Chou}}, \bibinfo {author} {\bibfnamefont {A.}~\bibnamefont {Collopy}},
  \bibinfo {author} {\bibfnamefont {C.}~\bibnamefont {Kurz}}, \bibinfo {author}
  {\bibfnamefont {Y.}~\bibnamefont {Lin}}, \bibinfo {author} {\bibfnamefont
  {M.}~\bibnamefont {Harding}}, \bibinfo {author} {\bibfnamefont
  {P.}~\bibnamefont {Plessow}}, \bibinfo {author} {\bibfnamefont
  {T.}~\bibnamefont {Fortier}}, \bibinfo {author} {\bibfnamefont
  {S.}~\bibnamefont {Diddams}}, \bibinfo {author} {\bibfnamefont
  {D.}~\bibnamefont {Leibfried}}, \ and\ \bibinfo {author} {\bibfnamefont
  {D.}~\bibnamefont {Leibrandt}},\ }\href@noop {} {\bibfield  {journal}
  {\bibinfo  {journal} {Science}\ }\textbf {\bibinfo {volume} {367}},\ \bibinfo
  {pages} {1458} (\bibinfo {year} {2020})}\BibitemShut {NoStop}%
\bibitem [{\citenamefont {Lin}\ \emph {et~al.}(2019)\citenamefont {Lin},
  \citenamefont {Leibrandt}, \citenamefont {Leibfried},\ and\ \citenamefont
  {Chou}}]{lin19}%
  \BibitemOpen
  \bibfield  {author} {\bibinfo {author} {\bibfnamefont {Y.}~\bibnamefont
  {Lin}}, \bibinfo {author} {\bibfnamefont {D.~R.}\ \bibnamefont {Leibrandt}},
  \bibinfo {author} {\bibfnamefont {D.}~\bibnamefont {Leibfried}}, \ and\
  \bibinfo {author} {\bibfnamefont {C.}~\bibnamefont {Chou}},\ }\href@noop {}
  {\bibfield  {journal} {\bibinfo  {journal} {arxiv: 1912.05866}\ } (\bibinfo
  {year} {2019})}\BibitemShut {NoStop}%
\bibitem [{\citenamefont {Meir}\ \emph {et~al.}(2019)\citenamefont {Meir},
  \citenamefont {Hegi}, \citenamefont {Najafian}, \citenamefont {Sinhal},\ and\
  \citenamefont {Willitsch}}]{meir19a}%
  \BibitemOpen
  \bibfield  {author} {\bibinfo {author} {\bibfnamefont {Z.}~\bibnamefont
  {Meir}}, \bibinfo {author} {\bibfnamefont {G.}~\bibnamefont {Hegi}}, \bibinfo
  {author} {\bibfnamefont {K.}~\bibnamefont {Najafian}}, \bibinfo {author}
  {\bibfnamefont {M.}~\bibnamefont {Sinhal}}, \ and\ \bibinfo {author}
  {\bibfnamefont {S.}~\bibnamefont {Willitsch}},\ }\href@noop {} {\bibfield
  {journal} {\bibinfo  {journal} {Faraday Discuss.}\ }\textbf {\bibinfo
  {volume} {217}},\ \bibinfo {pages} {561} (\bibinfo {year}
  {2019})}\BibitemShut {NoStop}%
\bibitem [{\citenamefont {Najafian}\ \emph {et~al.}(2020)\citenamefont
  {Najafian}, \citenamefont {Meir},\ and\ \citenamefont
  {Willitsch}}]{najafian20b}%
  \BibitemOpen
  \bibfield  {author} {\bibinfo {author} {\bibfnamefont {K.}~\bibnamefont
  {Najafian}}, \bibinfo {author} {\bibfnamefont {Z.}~\bibnamefont {Meir}}, \
  and\ \bibinfo {author} {\bibfnamefont {S.}~\bibnamefont {Willitsch}},\
  }\href@noop {} {\bibfield  {journal} {\bibinfo  {journal} {In preparation}\ }
  (\bibinfo {year} {2020})}\BibitemShut {NoStop}%
\bibitem [{\citenamefont {Biesheuvel}\ \emph {et~al.}(2016)\citenamefont
  {Biesheuvel}, \citenamefont {Karr}, \citenamefont {Hilico}, \citenamefont
  {Eikema}, \citenamefont {Ubachs},\ and\ \citenamefont
  {Koelemeij}}]{biesheuvel16a}%
  \BibitemOpen
  \bibfield  {author} {\bibinfo {author} {\bibfnamefont {J.}~\bibnamefont
  {Biesheuvel}}, \bibinfo {author} {\bibfnamefont {J.~P.}\ \bibnamefont
  {Karr}}, \bibinfo {author} {\bibfnamefont {L.}~\bibnamefont {Hilico}},
  \bibinfo {author} {\bibfnamefont {K.~S.~E.}\ \bibnamefont {Eikema}}, \bibinfo
  {author} {\bibfnamefont {W.}~\bibnamefont {Ubachs}}, \ and\ \bibinfo {author}
  {\bibfnamefont {J.~C.~J.}\ \bibnamefont {Koelemeij}},\ }\href@noop {}
  {\bibfield  {journal} {\bibinfo  {journal} {Nat. Commun.}\ }\textbf {\bibinfo
  {volume} {7}},\ \bibinfo {pages} {10385} (\bibinfo {year}
  {2016})}\BibitemShut {NoStop}%
\bibitem [{\citenamefont {Schiller}\ \emph {et~al.}(2014)\citenamefont
  {Schiller}, \citenamefont {Bakalov},\ and\ \citenamefont
  {Korobov}}]{schiller14a}%
  \BibitemOpen
  \bibfield  {author} {\bibinfo {author} {\bibfnamefont {S.}~\bibnamefont
  {Schiller}}, \bibinfo {author} {\bibfnamefont {D.}~\bibnamefont {Bakalov}}, \
  and\ \bibinfo {author} {\bibfnamefont {V.~I.}\ \bibnamefont {Korobov}},\
  }\href@noop {} {\bibfield  {journal} {\bibinfo  {journal} {{Phys. Rev.
  Lett.}}\ }\textbf {\bibinfo {volume} {113}},\ \bibinfo {pages} {023004}
  (\bibinfo {year} {2014})}\BibitemShut {NoStop}%
\bibitem [{\citenamefont {Schiller}\ and\ \citenamefont
  {Korobov}(2005)}]{schiller05a}%
  \BibitemOpen
  \bibfield  {author} {\bibinfo {author} {\bibfnamefont {S.}~\bibnamefont
  {Schiller}}\ and\ \bibinfo {author} {\bibfnamefont {V.}~\bibnamefont
  {Korobov}},\ }\href {2505} {\bibfield  {journal} {\bibinfo  {journal} {{Phys.
  Rev. A}}\ }\textbf {\bibinfo {volume} {71}},\ \bibinfo {pages} {032505}
  (\bibinfo {year} {2005})}\BibitemShut {NoStop}%
\bibitem [{\citenamefont {Jansen}\ \emph {et~al.}(2014)\citenamefont {Jansen},
  \citenamefont {Bethlem},\ and\ \citenamefont {Ubachs}}]{jansen14a}%
  \BibitemOpen
  \bibfield  {author} {\bibinfo {author} {\bibfnamefont {P.}~\bibnamefont
  {Jansen}}, \bibinfo {author} {\bibfnamefont {H.~L.}\ \bibnamefont {Bethlem}},
  \ and\ \bibinfo {author} {\bibfnamefont {W.}~\bibnamefont {Ubachs}},\
  }\href@noop {} {\bibfield  {journal} {\bibinfo  {journal} {J. Chem. Phys.}\
  }\textbf {\bibinfo {volume} {140}},\ \bibinfo {pages} {010901} (\bibinfo
  {year} {2014})}\BibitemShut {NoStop}%
\bibitem [{\citenamefont {Flambaum}\ and\ \citenamefont
  {Kozlov}(2007)}]{flambaum07a}%
  \BibitemOpen
  \bibfield  {author} {\bibinfo {author} {\bibfnamefont {V.}~\bibnamefont
  {Flambaum}}\ and\ \bibinfo {author} {\bibfnamefont {M.}~\bibnamefont
  {Kozlov}},\ }\href@noop {} {\bibfield  {journal} {\bibinfo  {journal} {Phys.
  Rev. Lett.}\ }\textbf {\bibinfo {volume} {99}},\ \bibinfo {pages} {150801}
  (\bibinfo {year} {2007})}\BibitemShut {NoStop}%
\bibitem [{\citenamefont {Kajita}\ \emph {et~al.}(2014)\citenamefont {Kajita},
  \citenamefont {Gopakumar}, \citenamefont {Abe}, \citenamefont {Hada},\ and\
  \citenamefont {Keller}}]{kajita14a}%
  \BibitemOpen
  \bibfield  {author} {\bibinfo {author} {\bibfnamefont {M.}~\bibnamefont
  {Kajita}}, \bibinfo {author} {\bibfnamefont {G.}~\bibnamefont {Gopakumar}},
  \bibinfo {author} {\bibfnamefont {M.}~\bibnamefont {Abe}}, \bibinfo {author}
  {\bibfnamefont {M.}~\bibnamefont {Hada}}, \ and\ \bibinfo {author}
  {\bibfnamefont {M.}~\bibnamefont {Keller}},\ }\href@noop {} {\bibfield
  {journal} {\bibinfo  {journal} {{Phys. Rev. A}}\ }\textbf {\bibinfo {volume}
  {89}},\ \bibinfo {pages} {032509} (\bibinfo {year} {2014})}\BibitemShut
  {NoStop}%
\bibitem [{\citenamefont {Kajita}(2016)}]{kajita16a}%
  \BibitemOpen
  \bibfield  {author} {\bibinfo {author} {\bibfnamefont {M.}~\bibnamefont
  {Kajita}},\ }\href@noop {} {\bibfield  {journal} {\bibinfo  {journal} {Appl.
  Phys. B}\ }\textbf {\bibinfo {volume} {122}},\ \bibinfo {pages} {203}
  (\bibinfo {year} {2016})}\BibitemShut {NoStop}%
\bibitem [{\citenamefont {Leibfried}(2012)}]{leibfried12a}%
  \BibitemOpen
  \bibfield  {author} {\bibinfo {author} {\bibfnamefont {D.}~\bibnamefont
  {Leibfried}},\ }\href@noop {} {\bibfield  {journal} {\bibinfo  {journal} {New
  J. Phys.}\ }\textbf {\bibinfo {volume} {14}},\ \bibinfo {pages} {023029}
  (\bibinfo {year} {2012})}\BibitemShut {NoStop}%
\bibitem [{\citenamefont {Bressel}\ \emph {et~al.}(2012)\citenamefont
  {Bressel}, \citenamefont {Borodin}, \citenamefont {Shen}, \citenamefont
  {Hansen}, \citenamefont {{I. Ernsting}},\ and\ \citenamefont
  {Schiller}}]{bressel12a}%
  \BibitemOpen
  \bibfield  {author} {\bibinfo {author} {\bibfnamefont {U.}~\bibnamefont
  {Bressel}}, \bibinfo {author} {\bibfnamefont {A.}~\bibnamefont {Borodin}},
  \bibinfo {author} {\bibfnamefont {J.}~\bibnamefont {Shen}}, \bibinfo {author}
  {\bibfnamefont {M.}~\bibnamefont {Hansen}}, \bibinfo {author} {\bibnamefont
  {{I. Ernsting}}}, \ and\ \bibinfo {author} {\bibfnamefont {S.}~\bibnamefont
  {Schiller}},\ }\href@noop {} {\bibfield  {journal} {\bibinfo  {journal}
  {Phys. Rev. Lett.}\ }\textbf {\bibinfo {volume} {108}},\ \bibinfo {pages}
  {183003} (\bibinfo {year} {2012})}\BibitemShut {NoStop}%
\bibitem [{\citenamefont {Morigi}\ and\ \citenamefont
  {Walther}(2001)}]{morigi01a}%
  \BibitemOpen
  \bibfield  {author} {\bibinfo {author} {\bibfnamefont {G.}~\bibnamefont
  {Morigi}}\ and\ \bibinfo {author} {\bibfnamefont {H.}~\bibnamefont
  {Walther}},\ }\href@noop {} {\bibfield  {journal} {\bibinfo  {journal} {Eur.
  Phys. J. D}\ }\textbf {\bibinfo {volume} {13}},\ \bibinfo {pages} {261}
  (\bibinfo {year} {2001})}\BibitemShut {NoStop}%
\bibitem [{\citenamefont {Home}(2013)}]{home13a}%
  \BibitemOpen
  \bibfield  {author} {\bibinfo {author} {\bibfnamefont {J.~P.}\ \bibnamefont
  {Home}},\ }\href@noop {} {\bibfield  {journal} {\bibinfo  {journal} {Adv. At.
  Mol. Opt. Phys.}\ }\textbf {\bibinfo {volume} {62}},\ \bibinfo {pages} {231}
  (\bibinfo {year} {2013})}\BibitemShut {NoStop}%
\bibitem [{\citenamefont {Wineland}\ \emph {et~al.}(1998)\citenamefont
  {Wineland}, \citenamefont {Monroe}, \citenamefont {Itano}, \citenamefont
  {Leibfried}, \citenamefont {King},\ and\ \citenamefont
  {Meekhof}}]{wineland98a}%
  \BibitemOpen
  \bibfield  {author} {\bibinfo {author} {\bibfnamefont {D.~J.}\ \bibnamefont
  {Wineland}}, \bibinfo {author} {\bibfnamefont {C.}~\bibnamefont {Monroe}},
  \bibinfo {author} {\bibfnamefont {W.~M.}\ \bibnamefont {Itano}}, \bibinfo
  {author} {\bibfnamefont {D.}~\bibnamefont {Leibfried}}, \bibinfo {author}
  {\bibfnamefont {B.~E.}\ \bibnamefont {King}}, \ and\ \bibinfo {author}
  {\bibfnamefont {D.~M.}\ \bibnamefont {Meekhof}},\ }\href@noop {} {\bibfield
  {journal} {\bibinfo  {journal} {J. Res. Natl. Inst. Stan.}\ }\textbf
  {\bibinfo {volume} {103}},\ \bibinfo {pages} {259} (\bibinfo {year}
  {1998})}\BibitemShut {NoStop}%
\bibitem [{\citenamefont {Meekhof}\ \emph {et~al.}(1996)\citenamefont
  {Meekhof}, \citenamefont {Monroe}, \citenamefont {King}, \citenamefont
  {Itano},\ and\ \citenamefont {Wineland}}]{meekhof96a}%
  \BibitemOpen
  \bibfield  {author} {\bibinfo {author} {\bibfnamefont {D.~M.}\ \bibnamefont
  {Meekhof}}, \bibinfo {author} {\bibfnamefont {C.}~\bibnamefont {Monroe}},
  \bibinfo {author} {\bibfnamefont {B.~E.}\ \bibnamefont {King}}, \bibinfo
  {author} {\bibfnamefont {W.~M.}\ \bibnamefont {Itano}}, \ and\ \bibinfo
  {author} {\bibfnamefont {D.~J.}\ \bibnamefont {Wineland}},\ }\href@noop {}
  {\bibfield  {journal} {\bibinfo  {journal} {Phys. Rev. Lett.}\ }\textbf
  {\bibinfo {volume} {76}},\ \bibinfo {pages} {1796} (\bibinfo {year}
  {1996})}\BibitemShut {NoStop}%
\bibitem [{\citenamefont {Leibfried}\ \emph {et~al.}(2003)\citenamefont
  {Leibfried}, \citenamefont {Blatt}, \citenamefont {Monroe},\ and\
  \citenamefont {Wineland}}]{leibfried03a}%
  \BibitemOpen
  \bibfield  {author} {\bibinfo {author} {\bibfnamefont {D.}~\bibnamefont
  {Leibfried}}, \bibinfo {author} {\bibfnamefont {R.}~\bibnamefont {Blatt}},
  \bibinfo {author} {\bibfnamefont {C.}~\bibnamefont {Monroe}}, \ and\ \bibinfo
  {author} {\bibfnamefont {D.}~\bibnamefont {Wineland}},\ }\href@noop {}
  {\bibfield  {journal} {\bibinfo  {journal} {Rev. Mod. Phys.}\ }\textbf
  {\bibinfo {volume} {75}},\ \bibinfo {pages} {281} (\bibinfo {year}
  {2003})}\BibitemShut {NoStop}%
\bibitem [{\citenamefont {Patterson}(2018)}]{patterson18}%
  \BibitemOpen
  \bibfield  {author} {\bibinfo {author} {\bibfnamefont {D.}~\bibnamefont
  {Patterson}},\ }\href@noop {} {\bibfield  {journal} {\bibinfo  {journal}
  {Phys. Rev. A}\ }\textbf {\bibinfo {volume} {97}},\ \bibinfo {pages} {033403}
  (\bibinfo {year} {2018})}\BibitemShut {NoStop}%
\bibitem [{\citenamefont {Tong}\ \emph {et~al.}(2010)\citenamefont {Tong},
  \citenamefont {Winney},\ and\ \citenamefont {Willitsch}}]{tong10a}%
  \BibitemOpen
  \bibfield  {author} {\bibinfo {author} {\bibfnamefont {X.}~\bibnamefont
  {Tong}}, \bibinfo {author} {\bibfnamefont {A.~H.}\ \bibnamefont {Winney}}, \
  and\ \bibinfo {author} {\bibfnamefont {S.}~\bibnamefont {Willitsch}},\
  }\href@noop {} {\bibfield  {journal} {\bibinfo  {journal} {{Phys. Rev.
  Lett.}}\ }\textbf {\bibinfo {volume} {105}},\ \bibinfo {pages} {143001}
  (\bibinfo {year} {2010})}\BibitemShut {NoStop}%
\bibitem [{\citenamefont {Tong}\ \emph {et~al.}(2012)\citenamefont {Tong},
  \citenamefont {Nagy}, \citenamefont {{\mbox{Yosa Reyes}}}, \citenamefont
  {Germann}, \citenamefont {Meuwly},\ and\ \citenamefont
  {Willitsch}}]{tong12a}%
  \BibitemOpen
  \bibfield  {author} {\bibinfo {author} {\bibfnamefont {X.}~\bibnamefont
  {Tong}}, \bibinfo {author} {\bibfnamefont {T.}~\bibnamefont {Nagy}}, \bibinfo
  {author} {\bibfnamefont {J.}~\bibnamefont {{\mbox{Yosa Reyes}}}}, \bibinfo
  {author} {\bibfnamefont {M.}~\bibnamefont {Germann}}, \bibinfo {author}
  {\bibfnamefont {M.}~\bibnamefont {Meuwly}}, \ and\ \bibinfo {author}
  {\bibfnamefont {S.}~\bibnamefont {Willitsch}},\ }\href@noop {} {\bibfield
  {journal} {\bibinfo  {journal} {Chem. Phys. Lett.}\ }\textbf {\bibinfo
  {volume} {547}},\ \bibinfo {pages} {1} (\bibinfo {year} {2012})}\BibitemShut
  {NoStop}%
\bibitem [{\citenamefont {Safronova}\ and\ \citenamefont
  {Safronova}(2011)}]{safronova11}%
  \BibitemOpen
  \bibfield  {author} {\bibinfo {author} {\bibfnamefont {M.~S.}\ \bibnamefont
  {Safronova}}\ and\ \bibinfo {author} {\bibfnamefont {U.~I.}\ \bibnamefont
  {Safronova}},\ }\href@noop {} {\bibfield  {journal} {\bibinfo  {journal}
  {Phys. Rev. A}\ }\textbf {\bibinfo {volume} {83}},\ \bibinfo {pages} {012503}
  (\bibinfo {year} {2011})}\BibitemShut {NoStop}%
\bibitem [{\citenamefont {Wu}\ \emph {et~al.}(2007)\citenamefont {Wu},
  \citenamefont {Ben}, \citenamefont {Li}, \citenamefont {Zheng}, \citenamefont
  {Chen},\ and\ \citenamefont {Yang}}]{wu07a}%
  \BibitemOpen
  \bibfield  {author} {\bibinfo {author} {\bibfnamefont {Y.-D.}\ \bibnamefont
  {Wu}}, \bibinfo {author} {\bibfnamefont {J.-W.}\ \bibnamefont {Ben}},
  \bibinfo {author} {\bibfnamefont {B.}~\bibnamefont {Li}}, \bibinfo {author}
  {\bibfnamefont {L.-J.}\ \bibnamefont {Zheng}}, \bibinfo {author}
  {\bibfnamefont {Y.-Q.}\ \bibnamefont {Chen}}, \ and\ \bibinfo {author}
  {\bibfnamefont {X.-H.}\ \bibnamefont {Yang}},\ }\href@noop {} {\bibfield
  {journal} {\bibinfo  {journal} {Chinese J. Chem. Phys.}\ }\textbf {\bibinfo
  {volume} {20}},\ \bibinfo {pages} {285} (\bibinfo {year} {2007})}\BibitemShut
  {NoStop}%
\bibitem [{\citenamefont {Zare}(1988)}]{zare88a}%
  \BibitemOpen
  \bibfield  {author} {\bibinfo {author} {\bibfnamefont {R.~N.}\ \bibnamefont
  {Zare}},\ }\href@noop {} {\emph {\bibinfo {title} {{Angular Momentum}}}}\
  (\bibinfo  {publisher} {John Wiley \& Sons},\ \bibinfo {address} {New York},\
  \bibinfo {year} {1988})\BibitemShut {NoStop}%
\bibitem [{\citenamefont {Langhoff}\ \emph {et~al.}(1987)\citenamefont
  {Langhoff}, \citenamefont {Bauschlicher~Jr},\ and\ \citenamefont
  {Partridge}}]{langhoff87}%
  \BibitemOpen
  \bibfield  {author} {\bibinfo {author} {\bibfnamefont {S.~R.}\ \bibnamefont
  {Langhoff}}, \bibinfo {author} {\bibfnamefont {C.~W.}\ \bibnamefont
  {Bauschlicher~Jr}}, \ and\ \bibinfo {author} {\bibfnamefont {H.}~\bibnamefont
  {Partridge}},\ }\href@noop {} {\bibfield  {journal} {\bibinfo  {journal} {J.
  Chem. Phys.}\ }\textbf {\bibinfo {volume} {87}},\ \bibinfo {pages} {4716}
  (\bibinfo {year} {1987})}\BibitemShut {NoStop}%
\bibitem [{\citenamefont {Bennett}(1970)}]{bennett70}%
  \BibitemOpen
  \bibfield  {author} {\bibinfo {author} {\bibfnamefont {R.}~\bibnamefont
  {Bennett}},\ }\href@noop {} {\bibfield  {journal} {\bibinfo  {journal} {Mon.
  Not. R. Astron. Soc.}\ }\textbf {\bibinfo {volume} {147}},\ \bibinfo {pages}
  {35} (\bibinfo {year} {1970})}\BibitemShut {NoStop}%
\bibitem [{\citenamefont {Germann}(2016)}]{germann16d}%
  \BibitemOpen
  \bibfield  {author} {\bibinfo {author} {\bibfnamefont {M.}~\bibnamefont
  {Germann}},\ }\href@noop {} {Ph.D. thesis},\ \bibinfo  {school} {University
  of Basel} (\bibinfo {year} {2016})\BibitemShut {NoStop}%
\bibitem [{\citenamefont {Gilmore}\ \emph {et~al.}(1992)\citenamefont
  {Gilmore}, \citenamefont {Laher},\ and\ \citenamefont {Espy}}]{gilmore92}%
  \BibitemOpen
  \bibfield  {author} {\bibinfo {author} {\bibfnamefont {F.~R.}\ \bibnamefont
  {Gilmore}}, \bibinfo {author} {\bibfnamefont {R.~R.}\ \bibnamefont {Laher}},
  \ and\ \bibinfo {author} {\bibfnamefont {P.~J.}\ \bibnamefont {Espy}},\
  }\href@noop {} {\bibfield  {journal} {\bibinfo  {journal} {J. Phys. Chem.
  Ref. Data}\ }\textbf {\bibinfo {volume} {21}},\ \bibinfo {pages} {1005}
  (\bibinfo {year} {1992})}\BibitemShut {NoStop}%
\bibitem [{\citenamefont {Frisch}\ \emph {et~al.}()\citenamefont {Frisch},
  \citenamefont {Trucks}, \citenamefont {Schlegel}, \citenamefont {Scuseria},
  \citenamefont {Robb}, \citenamefont {Cheeseman}, \citenamefont {Scalmani},
  \citenamefont {Barone}, \citenamefont {Mennucci}, \citenamefont {Petersson},
  \citenamefont {Nakatsuji}, \citenamefont {Caricato}, \citenamefont {Li},
  \citenamefont {Hratchian}, \citenamefont {Izmaylov}, \citenamefont {Bloino},
  \citenamefont {Zheng}, \citenamefont {Sonnenberg}, \citenamefont {Hada},
  \citenamefont {Ehara}, \citenamefont {Toyota}, \citenamefont {Fukuda},
  \citenamefont {Hasegawa}, \citenamefont {Ishida}, \citenamefont {Nakajima},
  \citenamefont {Honda}, \citenamefont {Kitao}, \citenamefont {Nakai},
  \citenamefont {Vreven}, \citenamefont {Montgomery}, \citenamefont {Peralta},
  \citenamefont {Ogliaro}, \citenamefont {Bearpark}, \citenamefont {Heyd},
  \citenamefont {Brothers}, \citenamefont {Kudin}, \citenamefont {Staroverov},
  \citenamefont {Kobayashi}, \citenamefont {Normand}, \citenamefont
  {Raghavachari}, \citenamefont {Rendell}, \citenamefont {Burant},
  \citenamefont {Iyengar}, \citenamefont {Tomasi}, \citenamefont {Cossi},
  \citenamefont {Rega}, \citenamefont {Millam}, \citenamefont {Klene},
  \citenamefont {Knox}, \citenamefont {Cross}, \citenamefont {Bakken},
  \citenamefont {Adamo}, \citenamefont {Jaramillo}, \citenamefont {Gomperts},
  \citenamefont {Stratmann}, \citenamefont {Yazyev}, \citenamefont {Austin},
  \citenamefont {Cammi}, \citenamefont {Pomelli}, \citenamefont {Ochterski},
  \citenamefont {Martin}, \citenamefont {Morokuma}, \citenamefont {Zakrzewski},
  \citenamefont {Voth}, \citenamefont {Salvador}, \citenamefont {Dannenberg},
  \citenamefont {Dapprich}, \citenamefont {Daniels}, \citenamefont {Farkas},
  \citenamefont {Foresman}, \citenamefont {Ortiz}, \citenamefont {Cioslowski},\
  and\ \citenamefont {Fox}}]{g09D01}%
  \BibitemOpen
  \bibfield  {author} {\bibinfo {author} {\bibfnamefont {M.~J.}\ \bibnamefont
  {Frisch}}, \bibinfo {author} {\bibfnamefont {G.~W.}\ \bibnamefont {Trucks}},
  \bibinfo {author} {\bibfnamefont {H.~B.}\ \bibnamefont {Schlegel}}, \bibinfo
  {author} {\bibfnamefont {G.~E.}\ \bibnamefont {Scuseria}}, \bibinfo {author}
  {\bibfnamefont {M.~A.}\ \bibnamefont {Robb}}, \bibinfo {author}
  {\bibfnamefont {J.~R.}\ \bibnamefont {Cheeseman}}, \bibinfo {author}
  {\bibfnamefont {G.}~\bibnamefont {Scalmani}}, \bibinfo {author}
  {\bibfnamefont {V.}~\bibnamefont {Barone}}, \bibinfo {author} {\bibfnamefont
  {B.}~\bibnamefont {Mennucci}}, \bibinfo {author} {\bibfnamefont {G.~A.}\
  \bibnamefont {Petersson}}, \bibinfo {author} {\bibfnamefont {H.}~\bibnamefont
  {Nakatsuji}}, \bibinfo {author} {\bibfnamefont {M.}~\bibnamefont {Caricato}},
  \bibinfo {author} {\bibfnamefont {X.}~\bibnamefont {Li}}, \bibinfo {author}
  {\bibfnamefont {H.~P.}\ \bibnamefont {Hratchian}}, \bibinfo {author}
  {\bibfnamefont {A.~F.}\ \bibnamefont {Izmaylov}}, \bibinfo {author}
  {\bibfnamefont {J.}~\bibnamefont {Bloino}}, \bibinfo {author} {\bibfnamefont
  {G.}~\bibnamefont {Zheng}}, \bibinfo {author} {\bibfnamefont {J.~L.}\
  \bibnamefont {Sonnenberg}}, \bibinfo {author} {\bibfnamefont
  {M.}~\bibnamefont {Hada}}, \bibinfo {author} {\bibfnamefont {M.}~\bibnamefont
  {Ehara}}, \bibinfo {author} {\bibfnamefont {K.}~\bibnamefont {Toyota}},
  \bibinfo {author} {\bibfnamefont {R.}~\bibnamefont {Fukuda}}, \bibinfo
  {author} {\bibfnamefont {J.}~\bibnamefont {Hasegawa}}, \bibinfo {author}
  {\bibfnamefont {M.}~\bibnamefont {Ishida}}, \bibinfo {author} {\bibfnamefont
  {T.}~\bibnamefont {Nakajima}}, \bibinfo {author} {\bibfnamefont
  {Y.}~\bibnamefont {Honda}}, \bibinfo {author} {\bibfnamefont
  {O.}~\bibnamefont {Kitao}}, \bibinfo {author} {\bibfnamefont
  {H.}~\bibnamefont {Nakai}}, \bibinfo {author} {\bibfnamefont
  {T.}~\bibnamefont {Vreven}}, \bibinfo {author} {\bibfnamefont {J.~A.}\
  \bibnamefont {Montgomery}, \bibfnamefont {{Jr.}}}, \bibinfo {author}
  {\bibfnamefont {J.~E.}\ \bibnamefont {Peralta}}, \bibinfo {author}
  {\bibfnamefont {F.}~\bibnamefont {Ogliaro}}, \bibinfo {author} {\bibfnamefont
  {M.}~\bibnamefont {Bearpark}}, \bibinfo {author} {\bibfnamefont {J.~J.}\
  \bibnamefont {Heyd}}, \bibinfo {author} {\bibfnamefont {E.}~\bibnamefont
  {Brothers}}, \bibinfo {author} {\bibfnamefont {K.~N.}\ \bibnamefont {Kudin}},
  \bibinfo {author} {\bibfnamefont {V.~N.}\ \bibnamefont {Staroverov}},
  \bibinfo {author} {\bibfnamefont {R.}~\bibnamefont {Kobayashi}}, \bibinfo
  {author} {\bibfnamefont {J.}~\bibnamefont {Normand}}, \bibinfo {author}
  {\bibfnamefont {K.}~\bibnamefont {Raghavachari}}, \bibinfo {author}
  {\bibfnamefont {A.}~\bibnamefont {Rendell}}, \bibinfo {author} {\bibfnamefont
  {J.~C.}\ \bibnamefont {Burant}}, \bibinfo {author} {\bibfnamefont {S.~S.}\
  \bibnamefont {Iyengar}}, \bibinfo {author} {\bibfnamefont {J.}~\bibnamefont
  {Tomasi}}, \bibinfo {author} {\bibfnamefont {M.}~\bibnamefont {Cossi}},
  \bibinfo {author} {\bibfnamefont {N.}~\bibnamefont {Rega}}, \bibinfo {author}
  {\bibfnamefont {J.~M.}\ \bibnamefont {Millam}}, \bibinfo {author}
  {\bibfnamefont {M.}~\bibnamefont {Klene}}, \bibinfo {author} {\bibfnamefont
  {J.~E.}\ \bibnamefont {Knox}}, \bibinfo {author} {\bibfnamefont {J.~B.}\
  \bibnamefont {Cross}}, \bibinfo {author} {\bibfnamefont {V.}~\bibnamefont
  {Bakken}}, \bibinfo {author} {\bibfnamefont {C.}~\bibnamefont {Adamo}},
  \bibinfo {author} {\bibfnamefont {J.}~\bibnamefont {Jaramillo}}, \bibinfo
  {author} {\bibfnamefont {R.}~\bibnamefont {Gomperts}}, \bibinfo {author}
  {\bibfnamefont {R.~E.}\ \bibnamefont {Stratmann}}, \bibinfo {author}
  {\bibfnamefont {O.}~\bibnamefont {Yazyev}}, \bibinfo {author} {\bibfnamefont
  {A.~J.}\ \bibnamefont {Austin}}, \bibinfo {author} {\bibfnamefont
  {R.}~\bibnamefont {Cammi}}, \bibinfo {author} {\bibfnamefont
  {C.}~\bibnamefont {Pomelli}}, \bibinfo {author} {\bibfnamefont {J.~W.}\
  \bibnamefont {Ochterski}}, \bibinfo {author} {\bibfnamefont {R.~L.}\
  \bibnamefont {Martin}}, \bibinfo {author} {\bibfnamefont {K.}~\bibnamefont
  {Morokuma}}, \bibinfo {author} {\bibfnamefont {V.~G.}\ \bibnamefont
  {Zakrzewski}}, \bibinfo {author} {\bibfnamefont {G.~A.}\ \bibnamefont
  {Voth}}, \bibinfo {author} {\bibfnamefont {P.}~\bibnamefont {Salvador}},
  \bibinfo {author} {\bibfnamefont {J.~J.}\ \bibnamefont {Dannenberg}},
  \bibinfo {author} {\bibfnamefont {S.}~\bibnamefont {Dapprich}}, \bibinfo
  {author} {\bibfnamefont {A.~D.}\ \bibnamefont {Daniels}}, \bibinfo {author}
  {\bibfnamefont {{\"O}.}~\bibnamefont {Farkas}}, \bibinfo {author}
  {\bibfnamefont {J.~B.}\ \bibnamefont {Foresman}}, \bibinfo {author}
  {\bibfnamefont {J.~V.}\ \bibnamefont {Ortiz}}, \bibinfo {author}
  {\bibfnamefont {J.}~\bibnamefont {Cioslowski}}, \ and\ \bibinfo {author}
  {\bibfnamefont {D.~J.}\ \bibnamefont {Fox}},\ }\href@noop {} {\enquote
  {\bibinfo {title} {Gaussian 09 {R}evision {D}.01},}\ }\bibinfo {note}
  {{G}aussian Inc., Wallingford CT, 2009}\BibitemShut {NoStop}%
\end{thebibliography}
\bibliographystyle{apsrev4-1}

\appendix*
\section*{Methods}

\subsection*{Experimental setup}

A molecular-beam machine was coupled to a linear radio-frequency ion trap in which $^{40}$Ca$^+$ AI and a $^{14}$N$_2^+$ MI were trapped simultaneously (Figure \ref{fig:cartoon} of the main text). A small magnetic field of 4.6 G defined the quantization axis along the viewing direction of the camera, orthogonal to the trap axis. A continuous-wave (CW) laser beam at 789 nm was split into two paths which were then superimposed on the trapping region in a counter-propagating configuration to form an optical lattice. The polarization of the beams was chosen parallel to the magnetic-field vector. The frequency difference, $\Delta f$, between the beam paths was matched to the frequency of the IP motional mode using acousto-optic modulators (AOM). Two pulsed dye-laser beams at 202 nm and 375 nm were used to produce $^{14}$N$_2^+$ ions by resonance-enhanced multi-photon ionization (REMPI) from the pulsed molecular beam of neutral $^{14}$N$_2$ neutral molecules \cite{tong10a,tong12a}. The ionization scheme was chosen to create $^{14}$N$_2^+$ in the lowest rotational levels of the $I = 0$ or $I = 2$ nuclear-spin isomers (corresponding to levels with even rotational angular momentum quantum numbers $N = 0, 2, 4,  \dots$) in the electronic and vibrational ground state. We achieved this isomeric selectivity by using a [$2+1'$] REMPI scheme consisting of two 202 nm photons at $49426$ cm$^{-1}$ on resonance with the S($0$) transition from the rovibronic ground-state ($N=0$) in neutral $^{14}$N$_2$. Therefore, the $I = 1$ isomer, with its lowest rotational state $N=1$ was excluded from the ionization. Two CW lasers at 397 nm and 866 nm were used for Doppler laser cooling of Ca$^+$, and another two laser beams at 729 nm and 854 nm were used for resolved-sideband cooling and coherent state manipulation.

Our experimental procedure was described in detail in Refs. \cite{meir19a,sinhal20}. Approximately ten Ca$^+$ ions were loaded into the trap and Doppler cooled on the $(4s)~^2S_{1/2}\leftrightarrow (4p)~^2P_{1/2}\leftrightarrow (3d)~^2D_{3/2}$ closed optical cycling transitions to form a string of Coulomb-crystallized ions. A single N$_2^+$ ion was then loaded into the trap using REMPI. The appearance of a dark ion in the string signalled the sympathetic cooling of a molecule. By lowering the trap depth, Ca$^+$ ions were successively ejected from the trap until a Ca$^+$ - N$_2^+$ two-ion string remained. The ions were cooled to the motional ground state of the IP motional mode by resolved sideband cooling on the $(3d)~^2D_{5/2}(m=-5/2)\leftarrow (4s)~^2S_{1/2}(m=-1/2)$ transitions of Ca$^+$ followed by quenching the $D_{5/2}$ state to the $S_{1/2}$ level through the $(4p)~^2P_{3/2}$ state and optical pumping back to the $S_{1/2}(m=-1/2)$ state. To prepare the Ca$^+$ in the metastable $D_{5/2}(m=-5/2)$ state, we used a $\pi$-pulse on the narrow $D_{5/2}(m=-5/2)\leftarrow S_{1/2}(m=-1/2)$ transition followed by a state-purification pulse \cite{sinhal20,chou17a}. The state-purification pulse projected the Ca$^+$ either to the $D_{5/2}(m=-5/2)$ or the $S_{1/2}$ state with a heralded signal (photon scattering) that allowed us to exclude experiments with improper state preparation. We chose the $D_{5/2}(m=-5/2)$ due to its reduced ac-Stark shift compared to the $S_{1/2}$ state which would have overwhelmed the molecular signal.

\subsection*{Phase-sensitive forces}
In Ref. \cite{sinhal20}, we used a similar type of force-spectroscopic method to the one presented here. In that work, we employed a smaller lattice-laser detuning from a specific, well known resonance in N$_2^+$ prepared in a specific quantum state and consequently required a smaller lattice-laser intensity and duration compared to the present study. Under these conditions, small effects such as the contribution of the ODF on Ca$^+$ to the motional excitation, the effects of far-detuned molecular resonances including other electronic transitions and the influence of the polarizability of the N$_2^+$ core electrons on the excitation strength could be neglected. 

For the objectives of the present work, i.e., the determination of an initially unknown quantum state of the molecule, larger detunings from molecular resonances and consequently higher laser intensity and duration were necessary. Here, we give a more complete account of phase-sensitive force spectroscopy 
that includes the effect of the lattice on both the atomic and the molecular.

\subsubsection*{Normal modes of the system}
The derivation of the normal modes of a Coulomb crystal of two ions with unequal masses in a harmonic trap is described in Refs. \cite{morigi01a,home13a}. Here, we only give the final results and definitions which will be used later in the extraction of the molecular ac-Stark shift from the experimental signal.

The present system is composed of two ions, a molecular ion with mass $m_1$=28~u and an atomic ion with mass $m_2$=40~u. The ions are confined in a harmonic potential characterized by a ``spring'' constant $u_0$ such that the frequency of a single particle is given by $\omega_i=\sqrt{u_0/m_i}$. Here, the subscripts $i=1,2$ refer to the molecular ion and the atomic ion, respectively.

When both ions are trapped together, they form a crystal due to the balance between their Coulomb repulsion, $F_c=-\frac{e^2/4\pi\varepsilon_0}{r_{12}^2}$, and the harmonic confinement of the trap. Here, $e$ is the electron charge, $\varepsilon_0$ is the vacuum permittivity and $r_{12}$ is the separation between the two ions. The distance between the equilibrium positions, $x_i^0$, of the ions in the crystal is given by,
\begin{equation}
    d=|x_2^0-x_1^0|=\sqrt[3]{\frac{e^2}{4\pi\varepsilon_0}\frac{2}{u_0}}.
\end{equation}
By measuring the trap-oscillation frequency of a single atomic ion, $\omega_2$, with relative uncertainty $\delta \omega_2/\omega_2 < 10^{-3}$, the distance $d$ can be calculated with a similar accuracy.

At the motional-excitation amplitudes reached in the present experiments, the ions performed only small oscillations around their equilibrium positions, $q_i=x_i-x_i^0$, such that it is safe to keep only harmonic terms in an expansion of the potential. The displacements of the individual particles, however, do not correspond to the normal modes of the system. To transform to a normal-mode basis, scaling and rotation transformations given by,
\begin{equation}\label{eq:transform}
    \begin{pmatrix}
        \beta_+ \\
        \beta_-
    \end{pmatrix}
    =
    \begin{pmatrix}
        \cos(\theta)/\sqrt{\mu} & -\sin(\theta) \\
        \sin(\theta)/\sqrt{\mu} & \cos(\theta)
    \end{pmatrix}
    \begin{pmatrix}
        q_1 \\
        q_2
    \end{pmatrix}.
\end{equation}
were performed. Here, $\beta_\pm$ are the in-phase $(-)$ and out-of-phase $(+)$ normal-mode displacements, $\theta$ is a rotation angle where $\tan(\theta)=1/\sqrt{\mu}-\sqrt{\mu}+\sqrt{1/\mu+\mu-1}$ and $\mu=m_2/m_1$. In the normal-mode picture, the system is equivalent to an ion of mass $m_2$ trapped in a 2D harmonic potential with two uncoupled mode frequencies,
\begin{equation}
    \Omega_\pm=\omega_2\sqrt{1+\mu\pm\sqrt{1+\mu^2-\mu}}.
\end{equation}
Note that a change of one unit in the mass of the molecular ion (as in the case of the chemical reaction described in Figure \ref{fig:N4sig} of the main text) corresponds to a relative change of the in-phase mode frequency of $\delta\Omega_-/\Omega_-\approx6\times10^{-3}$ which can readily be detected in the present experiments.   

\subsubsection*{Lattice excitation of the normal modes}
The optical lattice induces a modulated ac-Stark shift on both the atomic and molecular ion,
\begin{equation}\label{eq:lattice}
    \Delta E_i =2\Delta E_i^0 \left(1+\cos(2kq_i-\omega t+\phi_i^0)\right).
\end{equation}
Here, $\Delta E_i^0$ is the ac-Stark shift induced by a single lattice beam, $\omega=2\pi /\lambda$ is the lattice-laser angular frequency, $k=2\pi/\lambda$ is the lattice-laser $k$-vector and $\lambda\approx789$ nm is the lattice-laser wavelength. The phase, $\phi_i^0=2kx_i^0$, depends to the equilibrium positions of the particles. In the SP configuration, the phase difference between the two particles, $\phi_{21}^0=\phi_2^0-\phi_1^0=2kd_\textrm{SP}=2\pi n$ ($n$ an integer), is such that the phase of the lattice is equal for both particles. In the OP configuration, $\phi_{21}^0=2kd_\textrm{OP}=2\pi (n+1/2)$, and the particles experience opposite phases of the lattice (Figure \ref{fig:cartoon}b of the main text). An additional phase difference is attributed to the sign of the ac-Stark shift, $\pm|\Delta E_i^0|$, which depends the detuning of the lattice-laser frequency with respect to the dominant resonance in the relevant particle.

To get an intuition on how the ODF on the molecular ion and the atomic ion affect the excitation of the in-phase mode of the crystal, we first expand Eq. \ref{eq:lattice} in a Taylor series around the equilibrium positions and neglect constant terms of the potential which do not exert any force,
\begin{equation}
    \Delta E_i\approx-4k\Delta E_i^0q_i\sin(\omega t-\phi_i^0).
\end{equation}
In this approximation, the lattice exerts an oscillating spatially homogeneous force with an amplitude $F_i=4k\Delta E_i^0$ on particle $i$. In the case of a single particle, this force will lead to coherent excitation of motion in the trap. In the case of two ions with unequal masses, the ac-Stark shift on the in-phase mode can be obtained using Eq. \ref{eq:transform},
\begin{equation}
\begin{split}
    \Delta E_-&=-4k\left(\Delta E_1^0 \sqrt{\mu}\sin(\theta) \pm \Delta E_2^0\cos(\theta) \right) \beta_- \sin(\omega t)\\
    &\equiv-4k\Delta E_-^0\beta_-\sin(\omega t).
\end{split}
\end{equation}
Here, the $\pm$ sign corresponds to the SP and OP configurations, respectively, and $\Delta E_-^0$ defines the ``single-beam'' ac-Stark shift of the in-phase mode. Thus, the two forces exerted by the lattice on the two ions are combined to a single effective force on the in-phase mode with an amplitude, $F_-=4k\Delta E_-^0$. This force will lead to coherent excitation of the in-phase mode of the two-ion crystal. In this analytical derivation, higher order terms of the lattice potential were neglected which lead to squeezing of the motional states and to mixing of the in-phase and out-of-phase modes. In the data analysis, we used a classical simulation of the two-ion system to account for high-order terms in the lattice excitation \cite{sinhal20,meir19a}. 

By measuring the in-phase motional excitation amplitude both in the SP ($|\Delta E_-^0(\textrm{SP})|$) and in the OP configuration ($|\Delta E_-^0(\textrm{OP})|$), the amplitude of the ac-Stark shift, $|\Delta E_{m}^0|$, exerted by the lattice on the molecule can be determined,
\begin{equation}
    |\Delta E_{m}^0|\equiv|\Delta E_1^0|=\frac{||\Delta E_-^0(\textrm{SP})|+|\Delta E_-^0(\textrm{OP})||}{2\sqrt{\mu}\sin(\theta)},
\end{equation}
under the assumption that $|\Delta E_{m}^0|>|\Delta E_{a}^0|$. The phase of the ac-Stark shift is inferred from the relative amplitudes obtained in the SP and OP measurements assuming that the sign of the ac-Stark shift on the atomic ion is negative. 

\subsubsection*{ac-Stark shift of Ca$^+$}
The ac-Stark shift, $\Delta E^j$, of a level $j$ in Ca$^+$ is given by,
\begin{equation}
    \Delta E^j=-\frac{\alpha^j(\omega)}{2\varepsilon_0 c}I,
\label{eq:acshift}
\end{equation}
where $\varepsilon_0$ is the vacuum permittivity, $c$ is the speed of light, $I$ the laser intensity, and $\alpha^j(\omega)$ the dynamic polarizability of level $j$ which is given by,
\begin{equation}
    \alpha^j(\omega)=\alpha^j_s(\omega)+\left(\frac{3\cos^2\Theta-1}{2}\right)\frac{3m_j^2-J_j(J_j+1)}{J_j(2J_j-1)}\alpha^j_t(\omega).
\end{equation}
Here, $\alpha^j_s(\omega)$ and $\alpha^j_t(\omega)$ are the scalar and tensor dynamic polarizabilities of level $j$, $\Theta$ is the angle of linear polarization ($\Theta=0$ for $\pi$-polarized light) and $J_j$ and $m_j$ are the quantum numbers of the total angular momentum and its projection. 

The $(4s)~^2S_{1/2}$ state of Ca$^+$ has only a scalar contribution to the polarizability which results in $\alpha^S(\omega)=97.5$ au \cite{safronova11} for a lattice laser wavelength of 789.0 nm.

The $(3d)~^2D_{5/2}(m=-5/2)$ state of Ca$^+$ has both scalar and tensor contributions to the polarizability. With linearly $\pi$-polarized light, this state interacts only with high lying $F$ states. The contribution of these states to the polarizability decays rather slowly such that all states up to the continuum need to be taken into account \cite{safronova11}. The polarizability of this state results in, $\alpha^D(\omega)=4.44$ au \cite{safronova11} at a lattice-laser wavelength of 789.0 nm.

The low polarizability of the $(3d)~^2D_{5/2}(m=-5/2)$ state with linearly polarized light is comparable to the polarizability of the core electrons, $\alpha^{D,\textrm{core}}=3.03$ au \cite{safronova11} for this state. The core contribution is almost identical in the $(3d)~^2D_{5/2}$ and the $(4s)~^2S_{1/2}$ states ($\alpha^{S,\textrm{core}}=3.134$ au \cite{safronova11} in the $S$ state). Therefore, in a spectroscopic experiment on the $^2D_{5/2}(m=-5/2)\leftarrow ~^2S_{1/2}(m=-1/2)$ transition, the measured ac-Stark shift has contributions mostly from the polarizability of the valence electron in both states,
\begin{equation}
\begin{split}
    \Delta E^{D\leftarrow S} &= \left( \Delta E^{D} + \Delta E^{D,\textrm{core}} \right) - \left( \Delta E^{S} + \Delta E^{S,\textrm{core}} \right) \\
     & \approx \Delta E^{D} - \Delta E^{S}.
\end{split}
\label{eq:SDshift}
\end{equation}
However, in a lattice-excitation experiment, the contribution to the ac-Stark shift from the atomic ion is due to both the core and valence electrons in the relevant state,
\begin{equation}
    \Delta E_2^0\equiv\Delta E^{j,\textrm{lattice}} = \Delta E^{j} + \Delta E^{j,\textrm{core}}.
\label{eq:HeatShift}
\end{equation}

\subsubsection*{ac-Stark shift calibration}
The amplitude of the ac-Stark shift was determined from the Rabi-oscillation data (see, e.g., Figure \ref{fig:exsig} of the main text) by comparison to a calibration experiment. In this calibration, the Ca$^+$ - N$_2^+$ Rabi-oscillation signal was simulated by exciting motion on a two-ion Ca$^+$ - N$_2$H$^+$ string with a well-defined ac-Stark shift amplitude applied on the atomic ion \cite{sinhal20}. We used an N$_2$H$^+$ molecular ion for convenience due to its longer chemical lifetime. The mass difference of 1~u between N$_2^+$ and N$_2$H$^+$ amounts to only a small correction compared to other experimental errors. 

In the calibration experiment, Ca$^+$ was prepared in the $(4s)^2S_{1/2}\left(m=-1/2\right)$ state due to its large polarizability that allowed us to tune the desired ac-Stark shift. The ac-Stark shift was calibrated by a spectroscopic measurement on the $(3d)~^2D_{5/2}\left(m=-5/2\right) \leftarrow  ~^2S_{1/2}\left(m=-1/2\right)$ transition of Ca$^+$ and varied by changing the lattice-laser power. Six different ac-Stark shifts in the range $0.8 - 4.6$ kHz were applied for generating the calibration data. The spectroscopically measured shifts were the combined shifts of both spectroscopic levels (Eq. \ref{eq:SDshift}). However, since the lattice excitations were preformed in the $^2S_{1/2}\left(m=-1/2\right)$ state only, the $^2D_{5/2}\left(m=-5/2\right)$ contribution (Eq. \ref{eq:HeatShift}) was subtracted. 

\begin{figure}[t!]
	\centering
	\includegraphics[width=\linewidth,trim={0cm 0cm 0cm 0cm},clip]{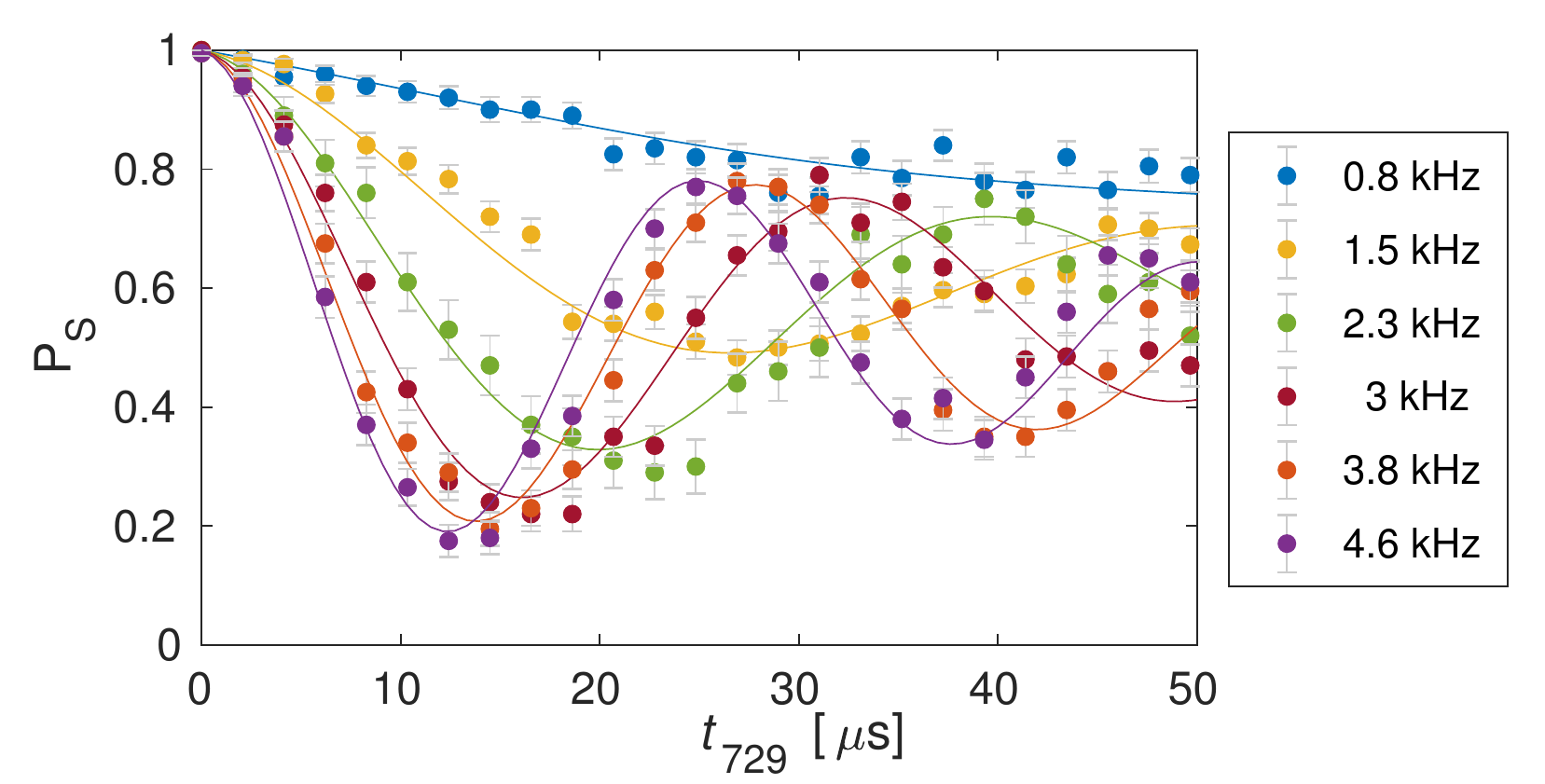}
 	\caption{Calibration measurements: Rabi-oscillation data from a two ion Ca$^+$ - N$_2$H$^+$ string. The Ca$^+$ ion is prepared in the $(4s)~^2S_{1/2}(m=-1/2)$ state with a well defined ac-Stark shift. The contribution of N$_2$H$^+$ to the excitation signal is corrected for iteratively (see text). The legend shows the second-iteration ac-Stark shift values including the contribution from N$_2$H$^+$}
	\label{fig:calibration}
\end{figure}

Subsequently, a lattice-excitation experiment was performed in which the ODF was applied for 3 ms with well-defined shifts on the Ca$^+$ - N$_2$H$^+$ system so that a Rabi-oscillation signal was obtained for each applied ac-Stark shift (see Figure \ref{fig:calibration}). We used these Rabi-oscillation signals as a reference for the ac-Stark shift amplitudes generated in the experiments described in the main text. In order to extract the amplitude from Rabi-oscillation data with unknown ac-Stark shift, an interpolation between the six calibrated ac-Stark shifts was performed using a single parameter (the ac-Stark shift) fitting function (see details in the SM of Ref. \cite{sinhal20}). 

Another advantage of using N$_2$H$^+$ as opposed to N$_2^+$ for the calibration experiment besides the longer lifetime is that it has no strong transitions near $789$ nm that could interfere with the calibration. There is, however, a residual force on the N$_2$H$^+$ due to highly detuned transitions in the molecule that must be taken into account (compare Figure 5b and the associated discussion in the main text). 
Since the ac-Stark shift on N$_2$H$^+$, $\Delta E_{\textrm{N}_2\textrm{H}^+}$ is small compared to the Ca$^+$ shift, $\Delta E_{\textrm{Ca}^+}$, in the $(4s)~^2S_{1/2}(m=-1/2)$ state, it was neglected to first order. This zero-order fitting function was then used to extract the ac-Stark shift of N$_2$H$^+$ (see e.g. Figure \ref{fig:N4sig}). An approximate value of $\Delta E_{\textrm{N}_2\textrm{H}^+} = - 0.81$ kHz was obtained corresponding to $\sim 15.8\%$ of $\Delta E_{\textrm{Ca}^+} = 5.41$ kHz measured with the same power. This shift was then added to the calibration data for a more realistic fitting function which now includes the total shift from both Ca$^+$ and N$_2$H$^+$. This procedure was iterated for a first order estimate of the ac-Stark shift of N$_2$H$^+$. The first iteration changed the measured shift to $\Delta E_{\textrm{N}_2\textrm{H}^+} = - 0.93$ kHz corresponding to $\sim 18.1\%$ of $\Delta E_{\textrm{Ca}^+}$. The second iteration yielded $\Delta E_{\textrm{N}_2\textrm{H}^+} = - 0.99$ kHz corresponding to $\sim 18.5\%$ of $\Delta E_{\textrm{Ca}^+}$. Thus, any residual error in the fitting function after the second iteration was neglected.

Figure \ref{fig:calibration} shows the results of the calibration experiments accounting for the effect of N$_2$H$^+$ excitation. The Ca$^+$ and N$_2$H$^+$ shifts were added together to produce the effective shift as indicated in the figure legend since the lattice was red detuned for both the atomic and molecular ions and the calibration was preformed in the SP$_\textrm{R}$ configuration (Figure \ref{fig:cartoon}b of the main text).

\subsubsection*{ac-Stark shift of N$_2^+$ including hyperfine structure}
The lattice-laser beam induced an ac-Stark shift on the molecular ion according to Eq. (\ref{eq:acshift}). This shift was calculated by summing up the contributions of the different transitions in the ro-vibrational band $A^2\Pi_u (v' = 2) \leftarrow X^2\Sigma_g^+ (v'' = 0)$,
\begin{equation}
    \alpha^j(\omega)=\sum_{k}\frac{2}{\hbar}\frac{\omega_{jk}}{\omega^2-{\omega_{jk}}^2}\left| \langle k | \bm{\mu} | j \rangle \right|^2,
\end{equation}
Here, $\hbar$ is the reduced Planck constant, $\omega_{jk}$ are the transition angular frequencies \cite{wu07a} and $|\langle k|\bm{\mu}|j\rangle|^2$ is the squared transition-dipole matrix element calculated using a spherical-tensor- algebra approach \cite{zare88a} using the value of the vibronic Einstein $A$ coefficient taken from Ref. \cite{langhoff87}. Here, the ground state $j$ is the $X^2\Sigma_g^+ (v'' = 0)$ vibronic state of N$_2^+$ which is adequately described by a Hund’s case (b) coupling scheme. The excited state $k$ is the $A^2\Pi_u (v' = 2)$ vibronic state described by an intermediate Hund’s case (a) / Hund's case (b) coupling scheme \cite{bennett70}. Hyperfine effects are included in the calculation of the dipole matrix element. The effect of mixing of states with different total angular momentum quantum number $J$ by hyperfine interactions \cite{germann16d} should only have a small effect on the ac-Stark shifts and was therefore not included in the calculations.  

Moreover, the contribution of additional vibrational bands in the $A^2\Pi_u (v' \neq 2)$ excited state and of the $B^2\Sigma_u^+ (v')$ electronic state were included. Here, due to the large detuning, only the rotationless transition frequencies were used \cite{gilmore92}. The corresponding Einstein $A$ coefficients were also taken from Ref. \cite{gilmore92}.

The core polarizability of N$_2^+$ was estimated from a calculation of the polarizability of N$_2^{2+}$. The calculation was performed using Gaussian 09 \cite{g09D01} at the CCSD/aug-cc-pVQZ theory level of theory. The value of the polarizability was found to be $\alpha^{\textrm{N}_2^+,\textrm{core}}\approx \alpha^{\textrm{N}_2^{2+}} =7.23$ au which corresponds to an ac-Stark shift of -390 Hz at our experimental parameters.

\end{document}